\documentclass{IEEEoj}
\usepackage{cite}
\usepackage{amsmath,amssymb,amsfonts}
\usepackage{algorithmic}
\usepackage{graphicx,color}
\usepackage{textcomp}
\usepackage{amsfonts}
\usepackage{bm}
\usepackage{tikz}
\usetikzlibrary{positioning, calc}
\usetikzlibrary{arrows.meta}
\usepackage{indentfirst}
\usepackage[ruled,vlined]{algorithm2e}
\usepackage{mathtools}
\usepackage{url}
\def\BibTeX{{\rm B\kern-.05em{\sc i\kern-.025em b}\kern-.08em
    T\kern-.1667em\lower.7ex\hbox{E}\kern-.125emX}}
\AtBeginDocument{\definecolor{ojcolor}{cmyk}{0.93,0.59,0.15,0.02}}
\def\OJlogo{\vspace{-4pt}\hskip-4pt\includegraphics[height=18pt]{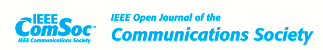}}
\begin{document}
\receiveddate{XX Month, XXXX}
\reviseddate{XX Month, XXXX}
\accepteddate{XX Month, XXXX}
\publisheddate{XX Month, XXXX}
\currentdate{11 January, 2024}
\doiinfo{OJCOMS.2024.011100}

\title{Score-Based Conditional Flow Models for MIMO Receiver Design with Superimposed Pilots}

\author{Ruhao Zhang\IEEEauthorrefmark{1}, Yupeng Li \IEEEauthorrefmark{2}, Yitong Liu \IEEEauthorrefmark{1}, Shijian Gao \IEEEauthorrefmark{3}, Jing Jin \IEEEauthorrefmark{2}, Hongwen Yang \IEEEauthorrefmark{1}, AND Jiangzhou Wang \IEEEauthorrefmark{4}\IEEEmembership{(Fellow, IEEE)}}
\affil{Beijing University of Posts and Telecommunications, Beijing, China}
\affil{China Mobile Research Institute, Beijing, China}
\affil{The Hong Kong University of Science and Technology (Guangzhou), China}
\affil{University of Kent, Canterbur, United Kingdom}
\corresp{CORRESPONDING AUTHOR: Yupeng Li (e-mail: liyupengtx@126.com).}
\markboth{Preparation of Papers for IEEE OPEN JOURNALS}{Author \textit{et al.}}

\begin{abstract}
Accurate channel state information (CSI) is vital for multiple-input multiple-output (MIMO) systems. However, superimposed pilots (SIP), which reduce overhead, introduce severe pilot contamination and data interference, complicating joint channel estimation and data detection. This paper proposes a conditional flow matching receiver (CFM-Rx), an unsupervised generative framework that learns directly from received signals, eliminating the need for labeled data and improving adaptability across diverse system settings. By leveraging flow-based generative modeling, CFM-Rx enables deterministic, low-latency inference and exploits model invertibility to capture the bidirectional nature of signal propagation. This framework unifies flow matching with score-based diffusion modeling via a moment-consistent ordinary differential equation (ODE), replacing stochastic differential equation (SDE) sampling with a deterministic and efficient process. Furthermore, it integrates receiver-side priors to ensure stable, data-consistent inference. Extensive simulation results across various MIMO configurations demonstrate that CFM-Rx consistently outperforms conventional estimators and state-of-the-art data-driven receivers, achieving notable gains in channel estimation accuracy and symbol detection robustness, particularly under severe pilot contamination.
\end{abstract}


\maketitle

\section{Introduction}

\IEEEPARstart{A}{ccurate} channel state information (CSI) is crucial for unlocking the full performance potential of multiple-input multiple-output (MIMO) systems, as it directly impacts signal detection, precoding, and system capacity \cite{mimo1, mimo2, chi2011,li2024channel}. Conventional CSI acquisition typically relies on orthogonal pilot sequences \cite{kim2008map, rag2005improving, bajwa2010com}, where pilot and data symbols are transmitted separately within each channel coherence block to avoid mutual interference. However, such orthogonal designs suffer from poor spectral efficiency, since dedicated resources must be reserved exclusively for pilot transmission. This inefficiency is particularly pronounced in massive MIMO or high-mobility scenarios, where pilot overhead becomes a major bottleneck to system throughput \cite{scaling, zilberstein2024diffusion, bjornson2017massive}.

To improve spectral efficiency, superimposed pilot (SIP) designs have attracted considerable attention as a promising alternative. By reusing pilots across spatial streams, users, or time-frequency resources, these schemes significantly reduce pilot overhead and improve overall spectral efficiency \cite{adhikary2013, xie2024cellfree, gupta2024affine, gu2024learning}. Nevertheless, this flexibility introduces new challenges: the reuse of SIPs leads to signal overlap in the received waveform, resulting in pilot contamination and data interference that degrade estimation and detection performance \cite{ashikhmin2012pilots, li2025learning, li2025iterative}. Effectively separating overlapping signals thus becomes a core difficulty in SIP receiver design.

Supervised deep learning (DL) approaches have shown strong performance where conventional model-driven estimators struggle, particularly under severe pilot contamination and inter-user interference. Unlike traditional receivers that separately implement channel estimation, equalization, and symbol detection based on analytical models, DL methods integrate these functions into a unified trainable network. In supervised training, the network takes received signals as inputs and the corresponding transmitted bit streams as labels, with parameters optimized via backpropagation. Once trained, it directly outputs bit streams from received signals during inference, thereby implicitly learning to mitigate interference and non-orthogonality by exploiting hidden structures in the data \cite{aoudia2021end,neumann2018learning, zou2025deeppilot, soltani2019deep, savaux2017lmmse}. This data-driven capability makes supervised DL receivers particularly attractive for complex environments where analytical models are difficult to construct.

However, despite their effectiveness, supervised DL methods face several practical limitations. As these models need to approximate complex nonlinear mappings, they often require large parameterized networks trained on massive labeled datasets, which are costly to obtain and store. Recent iterative AI-aided designs for SIP systems \cite{li2025learning} further rely on complex feedback loops between estimators and decoders, leading to high computational overhead and coupling between network architectures and specific scenarios. Moreover, their reliance on raw data to implicitly learn spatial channel characteristics leads to poor generalization across varying environments, antenna configurations, and pilot patterns. As a result, they often incur high retraining costs and limited scalability.

Motivated by these limitations, unsupervised learning has attracted considerable attention as a viable approach to reducing reliance on supervision. Unlike supervised methods that require labeled pilot–symbol pairs and fixed pilot configurations, the proposed approach exploits wireless channels directly, uncovering latent channel structures or signal representations that improve generalization across diverse system settings. This is particularly valuable in non-orthogonal communication systems, where collecting clean labeled training data is costly or impractical. Furthermore, unsupervised frameworks remove the constraint of fixed pilot layouts, enabling pilot-efficient and scalable receiver designs for in real-world deployment\cite{unsup2}.

In parallel, generative model techniques, notably diffusion models, have demonstrated strong potential for addressing inverse problems across vision, imaging, and wireless communications. Diffusion models, which define a stochastic denoising process to recover signals from noise, align naturally with the inverse nature of channel estimation and data detection tasks \cite{song2021score, ho2020denoising, song2019generative}. The authors of \cite{yudiffusion} propose a dual use of the diffusion model for CSI recovery. Several recent studies have extended diffusion models to jointly perform channel estimation and data detection under SIP conditions, yielding promising results \cite{zilberstein2024joint}.

However, diffusion models come with intrinsic drawbacks, including high sampling latency and limited determinism due to their iterative and stochastic nature. From a deployment perspective, modern wireless receivers are subject to stringent timing constraints for real-time processing. According to the 3GPP TS 38.214 \cite{214}, the PDSCH processing time for a high-capability UE is typically restricted to a sub-millisecond magnitude (e.g., approximately $0.1$ to $0.5$ ms depending on the sub-carrier spacing). While diffusion-based receivers, such as the one proposed in \cite{zilberstein2024joint}, typically require over 100 sampling steps to achieve convergence and upwards of 1,000 steps to reach optimal performance. Such excessive iterative complexity easily violates the strict latency budget and hardware throughput requirements of 5G NR and beyond. Moreover, the iterative nature of these models imposes a heavy burden on the computational power and on-chip memory bandwidth of mobile platforms, which are inherently limited by hardware cost and energy efficiency constraints. These limitations make them less suitable for real-time applications that demand low-latency and high-reliability inference, such as wireless receivers \cite{song2021score}. To address this, recent efforts have turned to flow-based generative models, especially flow matching (FM) techniques, which learn exact, invertible mappings between latent variables and observed signals \cite{lipman2023flow, ben2022matching, kobyzev2020normalizing}.

Unlike diffusion models, flow-based methods offer efficient, single-pass deterministic inference along with tractable and exact likelihood estimation, making them well-suited for high fidelity and strict latency applications \cite{papamakarios2019normalizing, kingma2018glow}. Furthermore, their invertibility aligns naturally with bidirectional signal processing tasks, such as joint channel estimation and data detection, where reversible mappings can capture the bidirectional nature of physical signal propagation \cite{zilberstein2024joint}. Although advanced frameworks like recursive flow \cite{jiang2026recursive} improve reconstruction accuracy, their complex recursive structures are primarily designed for channel estimation and remain computationally demanding for real-time joint processing. Despite their theoretical appeal, the application of flow models in wireless communication remains underexplored, and their performance under non-orthogonal settings is not well understood.

In this work, we propose a conditional flow matching receiver (CFM-Rx), a novel unsupervised framework for joint channel estimation and signal detection in MIMO systems with SIP designs. Unlike supervised learning approaches that rely on labeled training data to implicitly learn input-output mappings between received signals and transmitted bits, CFM-Rx directly models the underlying wireless channel features. From a Bayesian perspective, CFM-Rx can be viewed as a computational realization of Bayesian receiver design where the flow model serves as a data-driven prior distribution of the channel, and the inference process corresponds to sampling from the posterior distribution conditioned on the received signals. Built upon flow-based generative models, CFM-Rx formulates a moment-consistent ordinary differential equation (ODE) incorporating a conditional score function, which combines the prior-induced velocity fields with likelihood-based corrections, enabling deterministic and theoretically grounded inference at receiver inference. 
\footnote{The source code is available at: \url{https://github.com/fhghwericge/CFM-Rx-for-MIMO-Receiver-Design-with-Superimposed-Pilots}}

The key contributions of this paper are summarized as follows:
\begin{itemize}
\item 
We propose CFM-Rx, an unsupervised generative receiver framework for joint channel estimation and data detection under SIP configurations. Unlike supervised learning, which implicitly learns input–output mappings between received signals and transmitted bits, CFM-Rx directly models the underlying wireless channel environment. By integrating this learned model with real-time receiver-side information, such as received signal, pilot symbols, and coupling between channel and data matrices, CFM-Rx performs robust signal processing and effectively disentangles the non-orthogonal signal components through a physics-guided inference process. This design reduces the reliance on labeled training data, improves adaptability and generalization across diverse system settings, and enables robust demodulation under varying pilot patterns and channel conditions.

\item 
We propose a moment-consistent CFM framework that unifies ODE-based FM with score-based modeling, yielding a deterministic ODE formulation incorporating the score function. We derive a receiver-specific conditional velocity field that explicitly decomposes the inference trajectory into a prior-driven drift and a likelihood-based correction. This decomposition serves as a robust mechanism to mitigate error propagation by dynamically aligning the generative prior with real-time SIP measurements. The theoretical justification of the conditional score is provided in the Appendix. This integration ensures that the generated channel and data states remain physically consistent with the received signal, enabling accurate and stable joint estimation even under low signal-to-interference conditions.

\item
Extensive simulations demonstrate that CFM-Rx consistently outperforms conventional model-based and state-of-the-art data-driven methods in both channel estimation and symbol detection. Notably, ablation studies confirm that the proposed joint iterative refinement yields significant performance gains over cascaded estimation-equalization approaches, validating the necessity of the conditional velocity mechanism. The learned channel prior generalizes across modulation formats and unseen channel models, demonstrating intrinsic transferability. Furthermore, by eliminating the need for labeled training data, CFM-Rx significantly reduces dataset requirements while maintaining robustness and high precision across diverse system configurations.
\end{itemize}

\textit{Notation:} 
Scalars are denoted by italic letters (e.g., $x$), vectors by bold lowercase letters (e.g., $\mathbf{x}$), and matrices by bold uppercase letters (e.g., $\mathbf{X}$). The $i$-th element of $\mathbf{x}$ is $x_i$, and the $(i,j)$-th entry of $\mathbf{X}$ is $X_{i,j}$. The sets of real and complex numbers are denoted by $\mathbb{R}$ and $\mathbb{C}$, respectively. The expectation operator is $\mathbb{E}[\cdot]$, and $\|\mathbf{x}\|$ denotes the Euclidean norm. A circularly symmetric complex Gaussian distribution with mean $\mu$ and covariance $\sigma^2 \mathbf{I}$ is written as $\mathcal{N}(\mu,\sigma^2\mathbf{I})$. The Hadamard product is denoted by $\odot$, and the element-wise division is denoted by $\oslash$. Conditional distributions or densities are denoted as $p(x | y)$. A dot above a variable, e.g., $\dot{x}_t$, denotes its time derivative, and $\nabla_x  f(\mathbf{x})$ denotes the gradient of $f$ with respect to $\mathbf{x}$. $\nabla \cdot \mathbf{F}$ denotes the divergence of a vector field $\mathbf{F}$. $\partial_t p$ denotes the partial derivative of $p$ with respect to time $t$. For velocity fields and score functions in flow matching, we follow common notation and omit explicit conditioning on target variables, as these quantities are defined at the level of sample trajectories and their dependence is implicit. This choice is made solely for notational simplicity and does not affect the probabilistic interpretation.

\section{System Model}
We consider a typical 5G MIMO system with \(N_t\) transmit antennas and \(N_r\) receive antennas. In the frequency domain, there are \(N_S\) subcarriers, while in the time domain, the system operates on \(N_T\) consecutive orthogonal frequency division multiplexing (OFDM) symbols \cite{wang1, wang2}. After cyclic prefix removal and fast Fourier transform (FFT) at the receiver, the baseband input-output relation of the OFDM system is modeled as follows,
\begin{equation}
\mathbf{Y}_r = \sum_{l=1}^L \mathbf{H}_{r,l}\odot \mathbf{X}_l + \mathbf{N}_r,
\label{trans_model}
\end{equation}
where \(\mathbf{Y}_r \in \mathbb{C}^{N_S \times N_T}\) is the received signal matrix for the $r$-th antenna, which constitutes a slice of the complete received signal tensor \(\mathbf{Y} \in \mathbb{C}^{N_S \times N_T \times N_r}\). \(\mathbf{X}_l \in \mathbb{C}^{N_S \times N_T}\) is the matrix of symbols transmitted over the $l$-th spatial layer, derived from the symbol tensor \(\mathbf{X} \in \mathbb{C}^{N_S \times N_T \times L}\), where $L$ is the number of layers. The term \(\mathbf{H}_{r,l} \in \mathbb{C}^{N_S \times N_T}\) is the corresponding slice of the channel tensor \(\mathbf{H} \in \mathbb{C}^{N_S \times N_T \times N_r \times L}\). \(\mathbf{N}_r \in \mathbb{C}^{N_S \times N_T}\) is the additive white Gaussian noise matrix, which is independent and identically distributed according to \(\mathcal{N}(0,\sigma^2\mathbf{I})\).

In 5G new radio (NR) systems, demodulation reference signals (DMRS) are orthogonally allocated with data symbols, i.e., \(\mathbf{X}_l=\{\mathbf{D}_l,\mathbf{P}_l\}\) with pilots and data placed on disjoint resource elements (REs). While this orthogonal pilot (OP) scheme simplifies receiver processing, it comes at the cost of considerable pilot overhead, particularly in scenarios with large antenna arrays.

To provide a unified representation, we express the transmitted signal on each layer as
\begin{equation}
\mathbf{X}_l = \mathbf{M}_d \odot \mathbf{D}_l + \mathbf{M}_p \odot \mathbf{P}_l,\notag
\label{unified_model}
\end{equation}
where \(\mathbf{D}_l\) and \(\mathbf{P}_l\) denote the data and pilot symbol matrices, respectively, and \(\mathbf{M}_d, \mathbf{M}_p \in \mathbb{R}^{N_S \times N_T}\) are known weighting masks. Under the conventional OP scheme, \(\mathbf{M}_d\) and \(\mathbf{M}_p\) are binary-valued and complementary, such that pilots and data occupy disjoint REs.

In contrast to the OP scheme, superimposed pilot (SIP) transmission strategies have emerged as a viable approach for improving spectral efficiency. The core concept of SIP lies in avoiding dedicated pilot overhead by allowing data and pilot signals to be transmitted simultaneously over the same time-frequency resources. In this case, the transmitted signal reduces to
\[
\mathbf{X}_l = \sqrt{w}\,\mathbf{D}_l + \sqrt{v}\,\mathbf{P}_l,
\]
where \(w \in [0,1]\) and \(v=1-w\) are power allocation factors for the data and pilot matrices, respectively. This one-sided framework, favoring simplified model management, forms the foundation of the proposed scheme and enables the complete removal of the time-frequency resource overhead commonly incurred by orthogonal pilot transmission.

This superposition of pilots and data alters the signal structure, as both are non-orthogonally combined within the same resources. The resulting pilot--data interference substantially increases the difficulty of receiver processing and makes reliable demodulation more challenging. As a result, channel estimation and data detection become inherently coupled, motivating advanced receiver designs that jointly address both tasks.

\section{Generative Models and OT Path}

To address the ill-posed nature of SIP signal recovery, we leverage generative models to learn the complex channel distribution as a robust prior. Score-based diffusion and flow matching are unified through the probability flow ODE and optimal transport (OT), yielding smooth, constant-speed trajectories that are easy to learn and efficient to sample. This framework enables scalable, interpretable, and data-consistent modeling of complex MIMO channel and symbol distributions.

\subsection{FM via Interpolation ODE}

Score-based diffusion models typically employ a forward SDE to gradually perturb data into noise, and generate samples by reversing this stochastic process via a learned score function \cite{song2021score}. While generative models based on SDEs are powerful, their sampling process is often computationally intensive due to the stochastic nature of the path. To enable more efficient and direct sample generation, FM is proposed to directly regress a velocity field that drives a deterministic ODE, explicitly defining the continuous probability transport from the prior to the data distribution \cite{lipman2023flow}.
In FM, a continuous-time process $\{\mathbf{x}_t\}_{t\in[0,1]}$ is defined, where samples $\mathbf{x}_0$ are drawn from the target data distribution $p_0(\mathbf{x})$, and samples $\mathbf{x}_1$ are drawn from a known prior distribution $p_1(\mathbf{x})$, typically a standard Gaussian distribution. For \(t\in[0,1]\), the distribution of \(\mathbf{x}_t\) is defined as a linear interpolation between the two boundaries, that is,
\begin{equation}
\mathbf{x}_t = \alpha_t \mathbf{x}_0 + \sigma_t \mathbf{x}_1,\label{interpolation}
\end{equation}
where \(\{\alpha_t\}\) and \(\{\sigma_t\}\) are two interpolation coefficients. This process describes a continuous trajectory from a noise sample to a data sample. To characterize the dynamics of this path, we can compute its instantaneous velocity by taking the time derivative with respect to $t$. The function that governs this evolution is known as the velocity field, which is given by the following probability flow ODE:
\begin{equation}
v_t(\mathbf{x}_t) = \frac{\text{d}\mathbf{x}_t}{\text{d}t} = \dot\alpha_t \mathbf{x}_0 + \dot\sigma_t \mathbf{x}_1,\label{dynamic}
\end{equation}
where \(\dot\alpha_t = \text{d}\alpha_t/\text{d}t\) and \(\dot\sigma_t = \text{d}\sigma_t/\text{d}t\). For notational simplicity, we follow the convention in the flow-matching literature~\cite{lipman2023flow} and omit explicit conditioning on the target variable when referring to marginal probability flows or trajectory definitions, reserving conditional notation solely for probabilistic inference. Therefore, the dynamics of FM is represented by the probability flow ODE~\cite{lipman2023flow,chen2023probability,chen2018neural}. This formulation serves as the basis for constructing the learnable velocity field described below.

In practice, this velocity field is approximated by a neural network \(v_{\theta}(\mathbf{x}_t,t)\), which is trained by minimizing the following squared error objective:
\begin{equation}
\mathcal{L}_v(\theta) = \int_{0}^{1} 
\mathbb{E}\bigl[\|v_{\theta}(\mathbf{x}_t,t) - (\dot\alpha_t \mathbf{x}_0 + \dot\sigma_t \mathbf{x}_1)\|^2\bigr] \text{d}t.
\end{equation}

Given a sufficiently well trained $v_{\theta}$, one can sample the target \(\mathbf{x}_0\) by numerically solving Eq.~(\ref{dynamic}) in the reverse process as follows,
\begin{equation}
\dot{\mathbf{x}}_t = - v_t(\mathbf{x}_t),
\end{equation}
with initial condition \(\mathbf{x}_1 \sim \mathcal{N}(0, \text{I})\). This corresponds to the reverse process of FM. In practical inference, the continuous trajectory is discretized by uniformly partitioning the interval $[0,1]$ into $T$ steps.

\subsection{Optimal Transport for Efficient Flow Sampling}

The ODE formulation aligns naturally with OT, where generative paths can be interpreted as displacement interpolations~\cite{mccann1997convexity}. Specifically, OT theory seeks the most efficient transport plan that minimizes the cost of moving probability mass from the prior to the target data distribution. Under the OT framework with \(\alpha_t=1-t\) and \(\sigma_t=t\) for \(t\in[0,1]\), the derivatives are simplified to \(\dot\alpha_t=-1\) and \(\dot\sigma_t=1\). The velocity field is simplified to 
\[
v_t(\mathbf{x}_t | \mathbf{x}_0)=\dot\alpha_t \mathbf{x}_0 + \dot\sigma_t \mathbf{x}_1 = \mathbf{x}_1 - \mathbf{x}_0,
\]
which depends solely on the initial and terminal states.
These OT-based paths evolve linearly from noise to data under constant-speed flows, resulting in simple vector fields that are easier to approximate and more efficient for both training and sampling.



\begin{figure}[h]
\centering
\includegraphics[trim=14mm 270mm 80mm 245mm, clip, width=0.75\columnwidth]{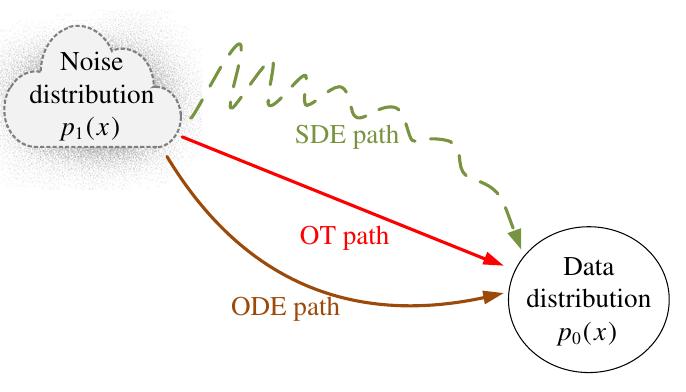}
\caption{Conceptual comparison of generative model trajectories.}
\label{fig:ode_vs_sde}
\end{figure}

Figure~\ref{fig:ode_vs_sde} illustrates the distinction between generative paths. Traditional SDE follows a noisy and erratic path from initial noise to the target data distribution. In contrast, the deterministic ODE path in FM provides a smooth and direct transition. Among these, the OT path, a special case of the ODE framework, corresponds to a straight-line, constant-velocity interpolation, yielding the simplest possible velocity field. This inherent simplicity contributes to both computational efficiency and stability.

This OT perspective, by simplifying the underlying vector fields to be learned, enables a more efficient and robust generative modeling approach. For wireless communications, this translates to a scalable and theoretically sound framework for learning the complex statistical structure of MIMO channels from unsupervised data. This capability is therefore particularly well-suited for challenging signal recovery tasks, such as MIMO detection and channel estimation. Although the generative framework is presented in a general form, in this work the flow model is instantiated to learn the wireless channel distribution only. The data symbols are inferred separately using known modulation priors and likelihood-based updates.

\section{Moment-Based CFM}

Adapting this general framework to receiver design, we derive a moment-consistent formulation that explicitly incorporates the received signal to guide sample generation. The moment-based framework unifies score-based diffusion and FM by reparameterizing the ODE with interpolation parameters $(\alpha_t, \sigma_t)$. Matching mean and variance ensures consistency with the flow path, while replacing the score with the posterior yields a deterministic conditional flow ODE that transports samples from a prior to the target conditional distribution.

\subsection{Moment Dynamics and Flow Derivation}
We reparameterize the score-based ODE using the FM parameters \(\alpha_t\) and \(\sigma_t\). The Fokker-Planck equation is a partial differential equation that describes the time evolution of the probability density function of the stochastic process defined by the SDE. By reformulating the Fokker–Planck equation governing the time evolution of the density under the forward SDE, a deterministic ODE can be identified whose solution induces identical density dynamics \cite[Eq.~(13)]{song2021score}. This leads to the score based ODE represented as follows
\begin{equation}
\text{d}\mathbf{x} = \left[ f(\mathbf{x},t) - \tfrac12g(t)^2\nabla_{\mathbf{x}}\log p_t(\mathbf{x})\right] \text{d}t. \label{scoreODE}
\end{equation}
This deterministic formulation is central to our work, providing an exact sampling path that we now parameterize. Specifically, by enforcing that the mean and variance of the distribution \(\mathbf{x}_t{|x_0} \sim \mathcal{N}\bigl(\alpha_t \mathbf{x}_0,\sigma_t^2 I\bigr)\) evolve consistently with the instantaneous drift and diffusion, we derive closed‐form expressions for \(f(\mathbf{x},t)\) and \(g(t)\). This moment-based analysis bridges the score-based diffusion perspective with the deterministic FM formulation in continuous time.

First, we determine the drift term. By assuming a linear form \(f(\mathbf{x}_t,t) = a(t)\mathbf{x}_t\) and equating the time derivative of the mean from the SDE dynamics with that from the FM interpolation (\(\mathbb{E}[\mathbf{x}_t{|x_0}]=\alpha_t \mathbf{x}_0\)), we solve for the drift coefficient, yielding \(a(t) = \dot{\alpha}_t / \alpha_t\). Thus, the drift term is given by \(f(\mathbf{x}_t,t) = (\dot{\alpha}_t/\alpha_t)\mathbf{x}_t\). The linear drift assumption governs the evolution of the latent channel variable along a tractable generative path, while the transmitted data symbols are treated separately through known modulation priors.

Next, we determine the diffusion coefficient \(g(t)\). The time derivative of the variance for a process with linear drift is given by \(\frac{d}{\text{d}t}\mathrm{Var}(\mathbf{x}_t) = 2a(t)\mathrm{Var}(\mathbf{x}_t) + g(t)^2\). By equating this with the time derivative of the known FM variance, \(\mathrm{Var}(\mathbf{x}_t)=\sigma_t^2\), and substituting our derived \(a(t)\), we can solve for the diffusion term, which gives \(g(t)^2 = 2\sigma_t(\dot\sigma_t - \frac{\dot\alpha_t}{\alpha_t}\sigma_t)\). For notational simplicity, we define \(\lambda_t = \dot\sigma_t - (\dot\alpha_t/\alpha_t)\sigma_t\), leading to \(g(t) = \sqrt{2\lambda_t\sigma_t}\).

Combining the expressions for the drift and diffusion coefficient and substituting them into \eqref{scoreODE}, the resulting flow ODE driven by the mean and variance dynamics can be expressed as
\begin{equation}
\frac{\text{d}\mathbf{x}}{\text{d}t}= \frac{\dot\alpha_t}{\alpha_t}\mathbf{x} -\lambda_t\sigma_t\nabla_{\mathbf{x}}\log p_t(\mathbf{x}).\label{scoreODE2}
\end{equation}

This deterministic formulation preserves both the mean and variance of the original SDE process and provides a principled basis for sample evolution in generative modeling. Distinct from existing interpolation-based velocity fields, it explicitly parameterizes the flow via the score function. This creates a direct interface to inject observation-dependent likelihoods, establishing the theoretical foundation for the conditional derivation that follows.

In the receiver signal processing, we are interested in sampling from the posterior conditional \(p(\mathbf{x} | \mathbf{z})\) rather than just \(p(\mathbf{x})\), as the receiver can leverage the observation, pilot signals, and so on. This motivates the justification of the conditional score function within the framework of (\ref{scoreODE2}). To enable conditional inference under observed context \(\mathbf{z}\), we define a conditional flow ODE driven by the score of the posterior distribution. The formal proof that this conditional velocity field induces a valid probability flow is provided in the Appendix. Specifically, at time \(t\) the conditional velocity field is given by
\begin{equation}
v_t(\mathbf{x}_t | \mathbf{z})
= \frac{\dot{\alpha}_t}{\alpha_t}\mathbf{x}_t
  - \lambda_t\sigma_t\nabla_{\mathbf{x}_t}\log p_t(\mathbf{x}_t| \mathbf{z}).\label{conditionalflowODE}
\end{equation}
Based on Liouville's theorem discussed in \cite{kardar2007statistical}, the conditional probability density satisfies the continuity equation driven by this velocity field, thereby ensuring the conservation of probability mass along the conditional trajectory.

\subsection{Framework for Conditional Flow Sampling}
The proposed framework consists of two conceptually distinct stages: unsupervised prior learning and observation-conditioned inference. In the learning stage, a flow-matching model is trained offline using unlabeled wireless channel realizations only, without any received signals, pilot matrices, or ground-truth labels. This procedure learns a scenario-dependent channel prior and does not constitute supervised learning. In the inference stage, the learned prior is conditioned on actual observations at the receiver through likelihood-based correction steps, guiding posterior sampling. This conditioning follows standard Bayesian principles and should be distinguished from supervised regression. Finally, data symbol inference is carried out analytically using known modulation priors and Tweedie's formula, reflecting the separate statistical roles of channel and symbols in the SIP model. The core of our approach is to jointly recover the channel matrix $\mathbf{H}$ and the data symbols $\mathbf{D}$ through an iterative process governed by CFM, where $\mathbf{H}$ is updated via conditional flows and $\mathbf{D}$ via analytic posterior inference. In this Bayesian formulation, the pre-trained generative network serves as the data-driven prior $p(\mathbf{H})$, while the flow matching inference process realizes sampling from the joint posterior $p(\mathbf{H}, \mathbf{D} | \mathbf{Y}, \mathbf{P})$ by fusing this prior with the likelihood.

Within the flow matching framework, the joint distribution of $(\mathbf{H}_t, \mathbf{D}_t)$ is Gaussian, satisfying the requisite conditions for defining these fields. By applying Bayes' theorem, each conditional velocity field can be decomposed into two interpretable components: an unconditional term that captures the prior dynamics, and a data-dependent correction term derived from the measurement likelihood. The resulting velocity fields for the channel $\mathbf{H}_t$ and the data symbols $\mathbf{D}_t$ are given by
\begin{subequations}
\begin{equation}
\begin{aligned}
&v_t(\mathbf{H}_t | \mathbf{D}_t, \mathbf{Y}, \mathbf{P}) =
\frac{\dot{\alpha}_t}{\alpha_t}\mathbf{H}_t - \lambda_t\sigma_t \nabla_{\mathbf{H}_t} \log p(\mathbf{H}_t | \mathbf{D}_t, \mathbf{Y}, \mathbf{P}) \\
&= \frac{\dot{\alpha}_t}{\alpha_t}\mathbf{H}_t - \lambda_t\sigma_t\bigl(\nabla_{\mathbf{H}_t}\log p(\mathbf{H}_t) + \nabla_{\mathbf{H}_t}\log p(\mathbf{Y} | \mathbf{H}_t,\mathbf{D}_t,\mathbf{P})\bigr)\\
&= v_t(\mathbf{H}_t) - \lambda_t\sigma_t\nabla_{\mathbf{H}_t}\log p(\mathbf{Y} | \mathbf{H}_t,\mathbf{D}_t,\mathbf{P}),
\end{aligned}
\label{vt_H}
\end{equation}
\begin{equation}
\begin{aligned}
&v_t(\mathbf{D}_t | \mathbf{H}_t, \mathbf{Y}, \mathbf{P}) =
\frac{\dot{\alpha}_t}{\alpha_t}\mathbf{D}_t - \lambda_t\sigma_t \nabla_{\mathbf{D}_t} \log p(\mathbf{D}_t | \mathbf{H}_t, \mathbf{Y}, \mathbf{P}) \\
&= \frac{\dot{\alpha}_t}{\alpha_t}\mathbf{D}_t - \lambda_t\sigma_t\bigl(\nabla_{\mathbf{D}_t}\log p(\mathbf{D}_t) + \nabla_{\mathbf{D}_t}\log p(\mathbf{Y} | \mathbf{D}_t,\mathbf{H}_t,\mathbf{P})\bigr)\\
&= v_t(\mathbf{D}_t) - \lambda_t\sigma_t\nabla_{\mathbf{D}_t}\log p(\mathbf{Y} | \mathbf{D}_t,\mathbf{H}_t,\mathbf{P}).
\end{aligned}
\label{vt_D}
\end{equation}
\end{subequations}

Here, $v_t(\mathbf{H}_t)$ and $v_t(\mathbf{D}_t)$ represent the unconditional velocity fields induced by the prior distribution, while the likelihood terms $\nabla\log p(\mathbf{Y}|\cdot)$ serve as correction scores. This decomposition provides a principled mechanism to combine a pretrained generative prior with real-time observation data.

As revealed by (\ref{vt_H}) and (\ref{vt_D}), the conditional velocity fields are governed by the evolving states of the channel and data matrices, while incorporating pilot symbols $\mathbf{P}$ and the received signal $\mathbf{Y}$ through the observation model. Within this framework, the unconditional velocity fields \(v_t(\mathbf{H}_t)\) and \(v_t(\mathbf{D}_t)\) provide the primary convergence guidance, whereas the likelihood-based correction refines the estimates by assimilating the prior information and the actual measurements. This complementary mechanism offers a principled means of integrating side information at the receiver, and mitigating the error propagation typical of cascaded designs, thereby laying the foundation for improved detection accuracy in subsequent analysis.

\section{Sampling via Conditional Flow Matching}
To translate the theoretical conditional flow into a practical algorithm, we employ a numerical sampling scheme that enforces data consistency. This section develops a conditional flow framework for joint channel and data detection, combining prior knowledge from flow matching with measurement-driven corrections. An iterative predictor–corrector scheme is then employed to produce posterior-consistent estimates. The proposed receiver follows a hybrid design: a learned flow-matching prior models channel statistics, while data detection is carried out analytically using constellation geometry and Bayesian updates. This decoupling enables scalability across modulation orders without retraining the flow model.

\subsection{Derivation of Velocity Components}
This subsection details the computation of the two constituent parts of the conditional velocity fields presented in (\ref{vt_H}) and (\ref{vt_D}), the unconditional velocity fields and the likelihood-based score functions.

The unconditional velocity field for the channel matrix, \(v_t(\mathbf{H}_t)\), is approximated by a neural network, denoted as \(v_\theta(\mathbf{H}_t, t)\), where \(\theta\) represents the network's trainable parameters. This network is designed as a conditional architecture that takes both the current channel state \(\mathbf{H}_t\) and the time step \(t\) as input. Specifically, it employs a U-Net like encoder-decoder structure to process the channel matrix. The time step \(t\) is first transformed into a high-dimensional embedding and then integrated into the network to condition the generation process, ensuring the output velocity is appropriate for the given time. This network is trained offline on a representative channel dataset. The complex-valued data is split into real and imaginary parts along the channel dimension before being fed into the model.

For the discrete target data symbols \(\mathbf{D}_0\) (which are to be inferred), the prior velocity \(v_t(\mathbf{D}_t)\) is computed analytically. It is defined by the general form
\[
v_t(\mathbf{D}_t) = \frac{\dot\alpha_t}{\alpha_t}\mathbf{D}_t - \lambda_t\sigma_t \nabla_{\mathbf{D}_t}\log p(\mathbf{D}_t),
\]
where the score term is handled using Tweedie's formula \cite{efron2011tweedie}, which connects the score to the minimum mean square error (MMSE) denoiser \(\mathbb{E}[\mathbf{D}_0|\mathbf{D}_t]\). This transformation allows us to bypass the intractable computation of the score function for a mixture distribution by expressing it in terms of the posterior expectation. This results in the expression:
\begin{equation}
v_t(\mathbf{D}_t) = \frac{\dot\alpha_t}{\alpha_t}\mathbf{D}_t - \lambda_t\sigma_t \frac{\alpha_t\mathbb{E}[\mathbf{D}_0| \mathbf{D}_t] - \mathbf{D}_t}{\sigma_t^2}. \label{vt_Dt}
\end{equation}
The conditional expectation is calculated by taking a weighted average over all constellation points \(\mathcal{X}\):
\begin{equation}
\mathbb{E}[\mathbf{D}_0 | \mathbf{D}_t]
= \frac{\displaystyle\sum_{x_k\in\mathcal X}
    x_k \exp\bigl(-\|\mathbf{D}_t - \alpha_t x_k\|^2/(2\sigma_t^2)\bigr)}
        {\displaystyle\sum_{x_k\in\mathcal X}
    \exp\bigl(-\|\mathbf{D}_t - \alpha_t x_k\|^2/(2\sigma_t^2)\bigr)},
\end{equation}
where \(k=\{1,2,...,2^M\}\), and \(M\) is the modulation order. Intuitively, this term acts as a soft-decision mechanism that aggregates information from all possible constellation points weighted by their Gaussian likelihoods, thereby guiding the noisy state $\mathbf{D}_t$ toward high-density regions corresponding to valid symbols.

To derive the likelihood scores \(\nabla\log p(\mathbf{Y}|\cdot)\), we start from the priors \({\mathbf{H}_t{|\mathbf{H}_0}} \sim \mathcal{N}(\mathbf{H}_0,\sigma_t^2 \mathbf{I}), \mathbf{D}_t{|\mathbf{D}_0} \sim \mathcal{N}(\mathbf{D}_0,\sigma_t^2 \mathbf{I}),\)
and the conditional observation model \(\mathbf{Y} \big| \mathbf{H}_0, \mathbf{D}_0, \mathbf{P} \sim \mathcal{N}\bigl(\mathbf{H}_0 \odot (w \mathbf{D}_0 + v \mathbf{P}), {\sigma}^2 \mathbf{I} \bigr),\) where \(\odot\) denotes element-wise multiplication. We may thus expand \(\mathbf{Y}\) in two equivalent ways:
\begin{align}
\mathbf{Y} &= \mathbf{H}_0\odot\bigl(w\mathbf{D}_0 + v\mathbf{P}\bigr) + {\mathbf{N}}\notag\\
&= \frac{\mathbf{H}_t}{\alpha_t}\odot\bigl(w\mathbf{D}_0 + v\mathbf{P}\bigr) - \frac{\sigma_t}{\alpha_t}\mathbf{H}_1\odot\bigl(w\mathbf{D}_0 + v\mathbf{P}\bigr) + {\mathbf{N}}\notag\\
&= \mathbf{H}_0\odot\bigl(w\frac{\mathbf{D}_t}{\alpha_t} + v\mathbf{P}\bigr) - w\frac{\sigma_t}{\alpha_t}\mathbf{H}_0\odot \mathbf{D}_1 + {\mathbf{N}},\notag
\end{align}
where \(\mathbf{H}_1, \mathbf{D}_1 \sim \mathcal{N}(\mathbf{0}, \mathbf{I})\) are independent standard Gaussian perturbations, and \({\mathbf{N}}\sim\mathcal{N}(\mathbf{0},{\sigma}^2\mathbf{I})\) is the observation noise. Consequently, the two conditional distributions follow:
\begin{subequations} 
\begin{align}
&\mathbf{Y} \bigl\lvert \mathbf{H}_t,\mathbf{D}_0,\mathbf{P} \sim \notag \\
&\mathcal{N}\bigl(
    \mathbf{H}_t \hspace{-0.1cm}\odot\hspace{-0.1cm} \tfrac{w\mathbf{D}_0 + v\mathbf{P}}{\alpha_t},
    \bigl(\tfrac{\sigma_t}{\alpha_t}\bigr)^{2}\hspace{-0.1cm}\mathrm{diag}\bigl(\lvert w\mathbf{D}_0 \hspace{-0.1cm}+\hspace{-0.1cm} v\mathbf{P}\rvert^2\bigr) + {\sigma}^2 \mathbf{I}
\bigr),
\end{align}
\begin{align}
&\mathbf{Y} \bigl\lvert \mathbf{H}_0,\mathbf{D}_t,\mathbf{P} \sim \notag \\
&\mathcal{N}\bigl(
    \mathbf{H}_0 \hspace{-0.1cm}\odot\hspace{-0.1cm} \bigl(w\tfrac{\mathbf{D}_t}{\alpha_t} + v\mathbf{P}\bigr),
    \bigl(w\tfrac{\sigma_t}{\alpha_t}\bigr)^{2}\hspace{-0.1cm}\mathrm{diag}\bigl(\lvert \mathbf{H}_0\rvert^2\bigr) + {\sigma}^2 \mathbf{I}
\bigr).
\end{align}
\end{subequations}
At each iteration, a coordinate-wise approach \cite{blei2017variational} is employed to turn the complex integrals into simple Gaussian partial forms, enabling closed-form partial derivatives with respect to \(\mathbf{H}_t\) and \(\mathbf{D}_t\). As a result, the conditional score function can be expressed as:
\begin{subequations}
\begin{equation}
\begin{aligned}
&\nabla_{\mathbf{H}_t}\log p(\mathbf{Y}|\mathbf{H}_t,\mathbf{D}_t,\mathbf{P})
\\&\approx
\frac{\bigl(w\mathbf{D}_t \hspace{-0.1cm}+\hspace{-0.1cm} v\mathbf{P}\bigr)^*/\alpha_t \odot\bigl(\mathbf{Y} \hspace{-0.1cm}- \hspace{-0.1cm}\mathbf{H}_t\hspace{-0.1cm}\odot\hspace{-0.1cm}\bigl((w\mathbf{D}_t+v\mathbf{P})/\alpha_t\bigr)\bigr)}
    {{\sigma}^2 + \bigl(\tfrac{\sigma_t}{\alpha_t}\bigr)^{2}\lvert w\mathbf{D}_t + v\mathbf{P}\rvert^2},
\end{aligned}
\label{score_Ht_wv}
\end{equation}
\begin{equation}
\begin{aligned}
&\nabla_{\mathbf{D}_t}\log p(\mathbf{Y}|\mathbf{H}_t,\mathbf{D}_t,\mathbf{P})
\\&\approx
\frac{w\mathbf{H}_t^*/\alpha_t \odot\bigl(\mathbf{Y} \hspace{-0.1cm}-\hspace{-0.1cm}\mathbf{H}_t\odot\bigl((w\mathbf{D}_t+v\mathbf{P})/\alpha_t\bigr)\bigr)}
    {{\sigma}^2 + \bigl(w\tfrac{\sigma_t}{\alpha_t}\bigr)^{2}\lvert \mathbf{H}_t\rvert^2},
\end{aligned}
\label{score_Dt_wv}
\end{equation}
\end{subequations}
where \(*\) denotes the conjugate transpose, and the terms \(\alpha_t\), \(\sigma_t\), and \(\lambda_t\) are the key hyperparameters governing the diffusion schedule. The complex-valued gradient is computed using the Wirtinger method. Stability of the likelihood gradients is ensured by bounding the denominator with \(\sigma^2\), and the step size \(c\) is adjusted based on the SNR to maintain numerical stability.
As established in the OT framework, a principled and efficient setting for these is \(\alpha_t=1-t\) and \(\sigma_t=t\). These expressions enable the alternating update of \(\mathbf{H}_t\) and \(\mathbf{D}_t\) by combining the pretrained flow matching drift with likelihood‐based score corrections.



\subsection{Iterative Inference via Predictor–Corrector Sampling}



The goal of the full inference procedure is to recover both the channel estimate and the detected data symbols by progressively transforming random noise into meaningful signal estimates. To achieve this, we adopt a reverse-time iterative framework that integrates both prior knowledge from the generative model and measurement consistency from the received signal. The process begins by initializing the latent channel and data variables, $\mathbf{H}_1$ and $\mathbf{D}_1$, with samples from a standard Gaussian distribution. Conditioned on the received signal $\mathbf{Y}$, the pilot matrix $\mathbf{P}$, and the pre-trained FM network $v_{\theta}$. Euler's method is used for numerical integration, discretizing the continuous process into \(T\) steps. These latent variables are refined through a backward loop of $T$ sampling steps.

A key challenge is that a single update step of (\ref{vt_H}) and (\ref{vt_D}) often cannot balance the generative prior with the measurement likelihood, leading to either unstable estimates or diminished guidance. To address this, we adopt an operator-splitting predictor–corrector scheme \cite{leimkuhler2015molecular}. By decoupling the update into two complementary stages, this strategy ensures that each iteration incorporates both sources of information. Moreover, multiple corrector refinements can be performed at each step, yielding a more robust and flexible trajectory that avoids overfitting. 

\begin{figure}[h]
\centering
\includegraphics[trim=60mm 28mm 60mm 23mm, clip, width=\columnwidth]{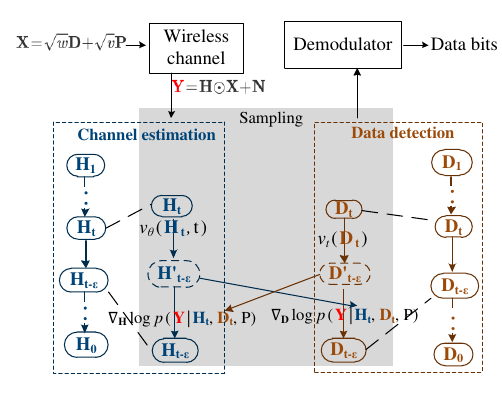}
\caption{Illustration of the conditional inference procedure with predictor-corrector iterations.}
\label{fig:inference_procedure}
\end{figure}

The predictor step corresponds to the unconditional update guided by the generative prior. Specifically, the reverse-time probability flow ODE is discretized using the Euler method \cite{hairer1993solving}, producing intermediate predictions $(\mathbf{H}'_{t-\epsilon}, \mathbf{D}'_{t-\epsilon})$:
\begin{subequations}
\begin{align}
    \mathbf{H}_{t - \epsilon}' &= \mathbf{H}_t - \epsilon \cdot v_\theta(\mathbf{H}_t, t), \\
    \mathbf{D}_{t - \epsilon}' &= \mathbf{D}_t - \epsilon \cdot v_t(\mathbf{D}_t),
\end{align}
\end{subequations}
where $(\mathbf{H}_t, \mathbf{D}_t)$ are the current state, $v_{\theta}(\mathbf{H}_t,t)$ is the learned network for the channel, and $v_t(\mathbf{D}_t)$ is the unconditional velocity fields for the data. $\epsilon=1/T$ is the sampling step size, such that evaluations are performed at $t \in \{0, \epsilon,2\epsilon,\ldots,1\}$. This step pushes the estimates toward a physically plausible state based purely on the prior knowledge.
The predictions $(\mathbf{H}'_{t-\epsilon}, \mathbf{D}'_{t-\epsilon})$ serve as an estimate at the subsequent time step $t-\epsilon$. 

The corrector step then enforces data consistency by incorporating the measurement likelihood. In practice, $K$ gradient ascent steps are applied to the log-likelihood, refining the predictor outputs:
\begin{subequations}
\begin{align}
\hspace{-0.2cm}\mathbf{H}_{t - \epsilon}' &\leftarrow \mathbf{H}_{t - \epsilon}' - c\cdot\epsilon \cdot \nabla_{\mathbf{H}_{t-\epsilon}'} \log p(\mathbf{Y} | \mathbf{H}_{t-\epsilon}', \mathbf{D}_{t-\epsilon}', \mathbf{P}), \\
\hspace{-0.2cm}\mathbf{D}_{t - \epsilon}' &\leftarrow \mathbf{D}_{t - \epsilon}' - c\cdot\epsilon \cdot \nabla_{\mathbf{D}_{t-\epsilon}'} \log p(\mathbf{Y} | \mathbf{H}_{t-\epsilon}', \mathbf{D}_{t-\epsilon}', \mathbf{P}).
\end{align}
\end{subequations}
Here, $c$ is the step size hyperparameter, which is typically set on the order of $1/K$, and $\nabla \log p(\mathbf{Y}|\cdot)$ is the likelihood score that steers the latent estimates toward consistency with the actual received signal $\mathbf{Y}$. The complete iterative inference pipeline, which combines predictor updates with likelihood-based corrections, is summarized in Algorithm \ref{alg:inference_procedure}.

\begin{algorithm}[h]
\caption{Predictor–corrector sampling.}
\label{alg:inference_procedure}
\small
\SetAlgoLined
\KwIn{$v_\theta$, $\mathbf{Y}$, $\mathbf{P}$, $T$, $K$, $c$}
Initialize $\mathbf{H}_1 \sim \mathcal{N}(0, \mathbf{I})$, $\mathbf{D}_{1} \sim \mathcal{N}(0, \mathbf{I})$, $\epsilon = 1/T$\;
\For{$i = T$ \KwTo $1$ }{
    $t = i/T$\;

    Compute $v_\theta(\mathbf{H}_t, t)$ from flow model\;
    Compute $v_t(\mathbf{D}_t)$ via constellation projection as in \eqref{vt_Dt}\;
    $\mathbf{H}_{t - \epsilon}' = \mathbf{H}_t - \epsilon \cdot v_\theta(\mathbf{H}_t, t)$\;
    $\mathbf{D}_{t - \epsilon}' = \mathbf{D}_t - \epsilon \cdot v_t(\mathbf{D}_t)$\;

    \For{$k = 1$ \KwTo $K$}{
        Compute $\nabla_{\mathbf{H}_{t-\epsilon}'} \log p(\mathbf{Y} | \mathbf{H}_{t-\epsilon}', \mathbf{D}_{t-\epsilon}', \mathbf{P})$ as in \eqref{score_Ht_wv}\;
        $\mathbf{H}_{t - \epsilon}' \leftarrow \mathbf{H}_{t - \epsilon}' - c\cdot\epsilon \cdot \nabla_{\mathbf{H}_{t-\epsilon}'} \log p(\mathbf{Y} | \mathbf{H}_{t-\epsilon}', \mathbf{D}_{t-\epsilon}', \mathbf{P})$\;

        Compute $\nabla_{\mathbf{D}_{t-\epsilon}'} \log p(\mathbf{Y} | \mathbf{H}_{t-\epsilon}', \mathbf{D}_{t-\epsilon}', \mathbf{P})$ as in \eqref{score_Dt_wv}\;
        $\mathbf{D}_{t - \epsilon}' \leftarrow \mathbf{D}_{t - \epsilon}' - c\cdot\epsilon \cdot \nabla_{\mathbf{D}_{t-\epsilon}'} \log p(\mathbf{Y} | \mathbf{H}_{t-\epsilon}', \mathbf{D}_{t-\epsilon}', \mathbf{P})$\;
    }
    $\mathbf{H}_{t - \epsilon} = \mathbf{H}_{t - \epsilon}'$\;
    $\mathbf{D}_{t - \epsilon} = \mathbf{D}_{t - \epsilon}'$\;
}
\KwOut{Recovered $\mathbf{H}_0$, $\mathbf{D}_{0}$}
\end{algorithm}

Through this iterative coupling, the evolving states achieve a principled integration of prior knowledge, observational evidence, and symbolic constraints, enabling a coherent and interpretable progression toward the true conditional posterior. This alternating structure is superior to a single unified update as it explicitly decouples the prior-driven evolution from the measurement-driven refinement. By isolating these operations, the framework circumvents numerical instabilities and error accumulation that typically occur during the simultaneous optimization of coupled channel and data variables. Furthermore, the $K$ corrector steps provide a fine-grained mechanism to robustly guide the sampling trajectory, which is vital for accurately disentangling the strong interference between channel and data matrices inherent in SIP scenarios. The proposed framework incorporates several synergistic components that collectively elevate both accuracy and robustness: 
\begin{itemize}
\item 
Generative prior dynamics: The velocity field for $\mathbf{H}_t$ is driven by a pretrained flow-matching model, effectively capturing structural patterns intrinsic to channel data and guiding its unconditional evolution with learned domain priors. In contrast, the velocity field for $\mathbf{D}_t$ is computed analytically using Tweedie’s identity, ensuring precise updates grounded in the current distribution and measurement model.
\item 
Diffusion scheduling: The evolution of $\mathbf{H}_t$ and $\mathbf{D}_t$ is regularized through scheduling parameters $\alpha_t$ and $\sigma_t$, which regularize the sampling trajectories to ensure stable and efficient convergence.
\item 
Observation likelihood: The measurement model contributes a conditional score as a data-driven correction term, adaptively steering the updates toward regions consistent with the posterior.
\item 
Symbolic constraints: The discrete constellation prior on $\mathbf{D}_t$ is explicitly incorporated, improving the physical plausibility of the generated samples.
\end{itemize}

By tightly integrating these components, the algorithm transcends conventional estimation pipelines, achieving data-consistent, physically interpretable results while leveraging both generative priors and analytical tractability.

\section{Simulation Results}

\subsection{Simulation Setup}
We consider a MIMO system with $N_t=4$ transmit and $N_r=4$ receive antennas. The wireless channel is simulated based on the 3GPP clustered delay line (CDL) channel model, as specified in TR 38.901\cite{901}. We configure the simulation with a carrier frequency of $3.5$\,GHz, a delay spread of $300$\,ns, and a mobility speed of $3$\,km/h. Additionally, the duration for each OFDM slot is \( T_{\text{slot}} = 1 \, \text{ms} \). The channel is realized independently across different frames.

Each data frame comprises 12 physical downlink shared channel (PDSCH) symbols and 48 subcarriers, with the transmitted symbols modulated using quadrature amplitude modulation (QAM). System performance is evaluated over a SNR range from $-10$\,dB to $20$\,dB in $5$\,dB intervals, where $\mathrm{SNR} = 10\log_{10}(\mathbb{E}[\|\mathbf{X}\|^2]/\mathbb{E}[\|\mathbf{N}\|^2])$. To ensure a fair comparison, the total transmit power $\mathbb{E}[\|\mathbf{X}\|^2]$ is normalized and kept constant across schemes. Under this setting, we investigate two representative pilot configurations:
\begin{itemize}
  \item \textbf{SIP:} Following the transmission model in \eqref{trans_model}, power allocation factors are set to $w=0.9$ for data symbols and $v=0.1$ for pilot symbols.
  \item \textbf{OP:} Pilots are placed on dedicated time-frequency REs in a comb-type pattern. 
\end{itemize}
In both cases, the pilot sequence is generated by mapping random bits onto quadrature phase shift keying (QPSK) symbols and is assumed to be perfectly known at the receiver.

\begin{figure}[h]
    \centering
    \includegraphics[trim=6mm 4mm 6mm 6mm, clip, width=\columnwidth]{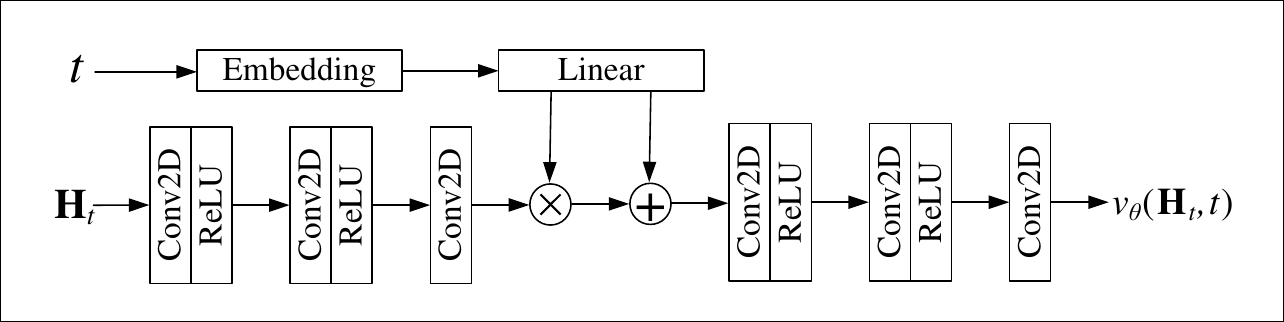}
    \caption{The architecture of the conditional U-Net used to approximate the velocity field $v_\theta(H_t, t)$.}
    \label{fig:network_arch}
\end{figure}

The proposed generative model is driven by the velocity field network, $v_\theta(H_t, t)$, which is implemented as a U-Net architecture with a 3-layer encoder and a 3-layer decoder, as illustrated in Fig.~\ref{fig:network_arch}. Training is performed using the Adam optimizer with an initial learning rate of $5 \times 10^{-4}$, decayed via a cosine annealing schedule. A batch size of 256 is used over 1000 epochs. The dataset is partitioned into training, validation, and testing sets with an 8:1:1 ratio. For the inference procedure, we numerically integrate the reverse-time conditional ODE using the iterative predictor-corrector scheme detailed in Algorithm~\ref{alg:inference_procedure}. The integration is performed over $T=30$ discrete time steps. Within each time step, the corrector stage consists of $K=5$ refinement steps, each utilizing a learning rate of $c = 1/K$ to update the estimates based on the measurement likelihood.


\subsection{Benchmarks}
The proposed conditional flow matching for receiver (CFM-Rx) is evaluated against the following baselines, with all methods tested under both the SIP and OP pilot configurations:
\begin{itemize}
  \item \textbf{5G-NR}: A modular receiver using LMMSE channel estimation followed by LMMSE equalization.
  \item \textbf{DM-JED}: A state-of-the-art diffusion model for joint channel estimation and data detection \cite{zilberstein2024joint}, trained in an unsupervised manner on wireless channel data.
  \item \textbf{E2E}: A fully convolutional neural network for end-to-end supervised learning \cite{aoudia2021end}, mapping received signals directly to the ground-truth data bits.
\end{itemize}

For the 5G-NR, the LMMSE estimator \cite{soltani2019deep, savaux2017lmmse} is expressed as followsyy:
\begin{equation}
\hat{\mathbf{h}}^{\mathrm{LMMSE}}=\mathbf{R}_{\mathbf{h}\mathbf{h}_p}\big(\mathbf{R}_{\mathbf{h}_p\mathbf{h}_p}+\sigma^2\mathbf{I}\big)^{-1}\hat{\mathbf{h}}_{p}^{\mathrm{LS}},
\end{equation}
where $\mathbf{R}_{\mathbf{h}\mathbf{h}_p}$ and $\mathbf{R}_{\mathbf{h}_p\mathbf{h}_p}$ are channel correlation matrices, $\hat{\mathbf{h}}_{p}^{\mathrm{LS}}$ is the results of least square estimates. Two variants of the LMMSE estimator are considered regarding these matrices: the \emph{orcale} case, assuming perfect knowledge of the channel covariance matrices (suffix \textbf{-O})\cite{savaux2017lmmse}, and the \emph{practical} case, where the covariance matrices are computed statistically from a set of channel snapshots (suffix -P)\cite{noh2006low}. Data symbols are then recovered via LMMSE equalization \cite{jiang2011performance} as $\hat{\mathbf{X}}=(\hat{\mathbf{H}}^H\hat{\mathbf{H}}+\sigma^2\mathbf{I})^{-1}\hat{\mathbf{H}}^H \mathbf{Y}$.

For DM-JED, a state-of-the-art unsupervised generative model built on a score-based SDE, enables direct comparison with our proposed CFM-Rx framework. The E2E model exemplifies the fully supervised paradigm, typically requiring complex architectures and large labeled datasets, with limited generalization and frequent retraining for configuration changes such as pilot patterns or modulation schemes.

\subsection{Channel Estimation and Link-Level Performance}

\begin{figure}[h]
  \centering
  \includegraphics[width=3.5in,height=2.7in]{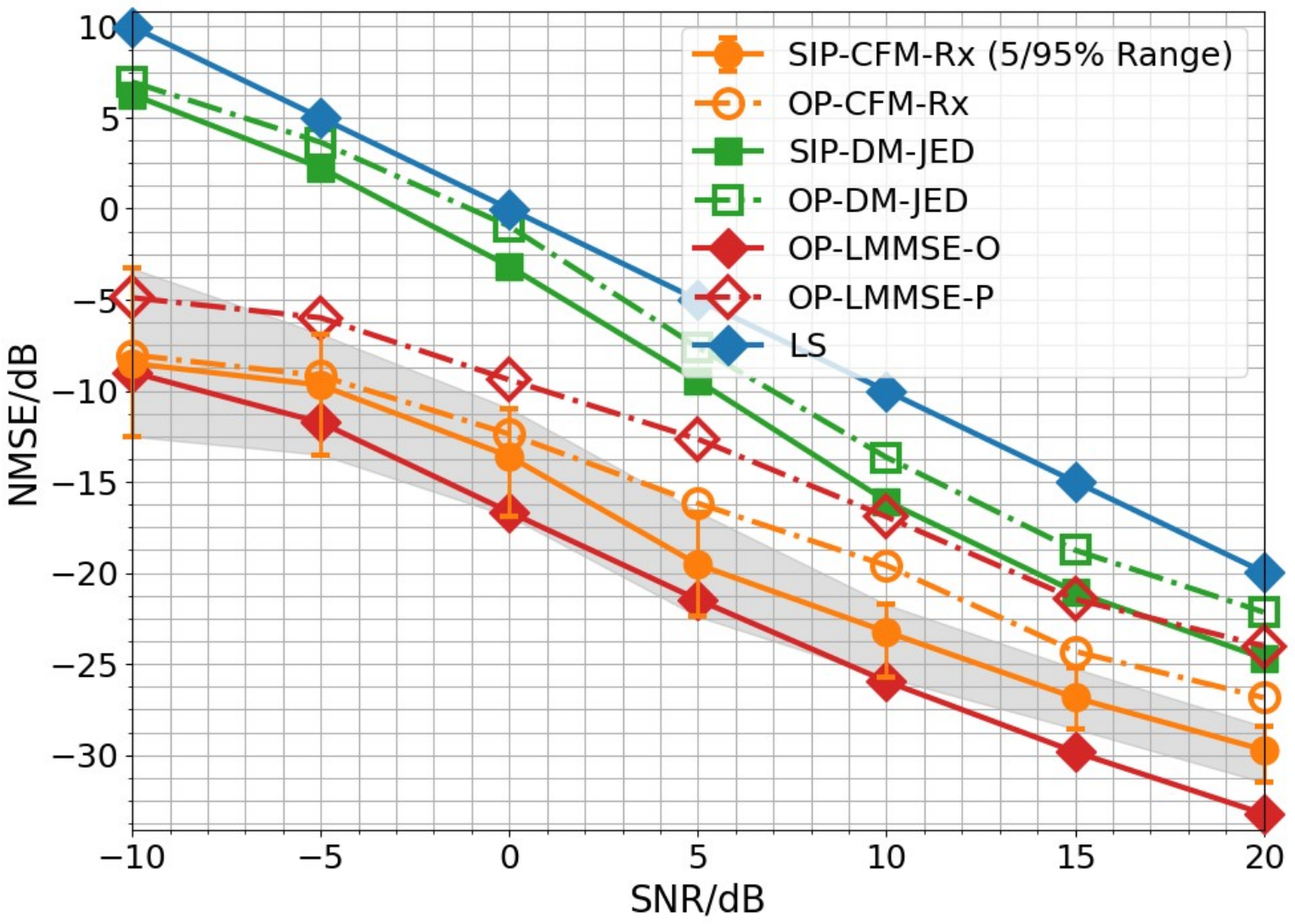}
  \caption{The performance comparisons of different channel estimation schemes.}
  \label{fig:NMSE}
\end{figure}

Figure~\ref{fig:NMSE} illustrates the channel estimation performance in terms of NMSE versus SNR. The orcale LMMSE estimator represents the theoretical performance ceiling, while the LMMSE and LS estimators serve as fundamental performance baselines. The proposed CFM-Rx is compared with the DM-JED estimator. Both data-driven methods are evaluated under OP and SIP pilot configurations.

It can be seen from the figure that CFM-Rx demonstrates consistently favorable performance across the entire SNR range. Under the SIP scheme, it outperforms DM-JED by 5 to 7~dB in NMSE at $\text{SNR} \ge 10$~dB. At low SNRs, the gap widens substantially, surpassing 10~dB for SNRs below 5~dB and reaching nearly 15~dB at -10~dB. The 90\% confidence interval ranges approximately 2 dB at high SNR and nearly 10 dB at low SNR, indicating that the proposed scheme exhibits better convergence at higher SNRs. The LS estimator exhibits the highest NMSE, confirming its role as a benchmark lower bound.

With orcale LMMSE serving as the performance ceiling, CFM-Rx with SIP closely approaches this bound across the entire SNR range. Compared with the practical LMMSE estimator, which shows noticeable degradation relative to its orcale counterpart, CFM-Rx maintains a significant advantage, particularly in the low-to-medium SNR regime. For both data-driven schemes, SIP consistently outperforms OP under the considered simulation settings. This performance gain arises because SIP distributes pilot and data power over all REs, providing denser signal information for learning and enabling more accurate channel reconstruction. The proximity of CFM-Rx to the LMMSE bound indicates that the learned generative flow effectively captures the underlying channel prior, allowing robust estimation even under noisy conditions.

\begin{figure}[h]
  \centering
  \includegraphics[width=3.5in,height=2.7in]{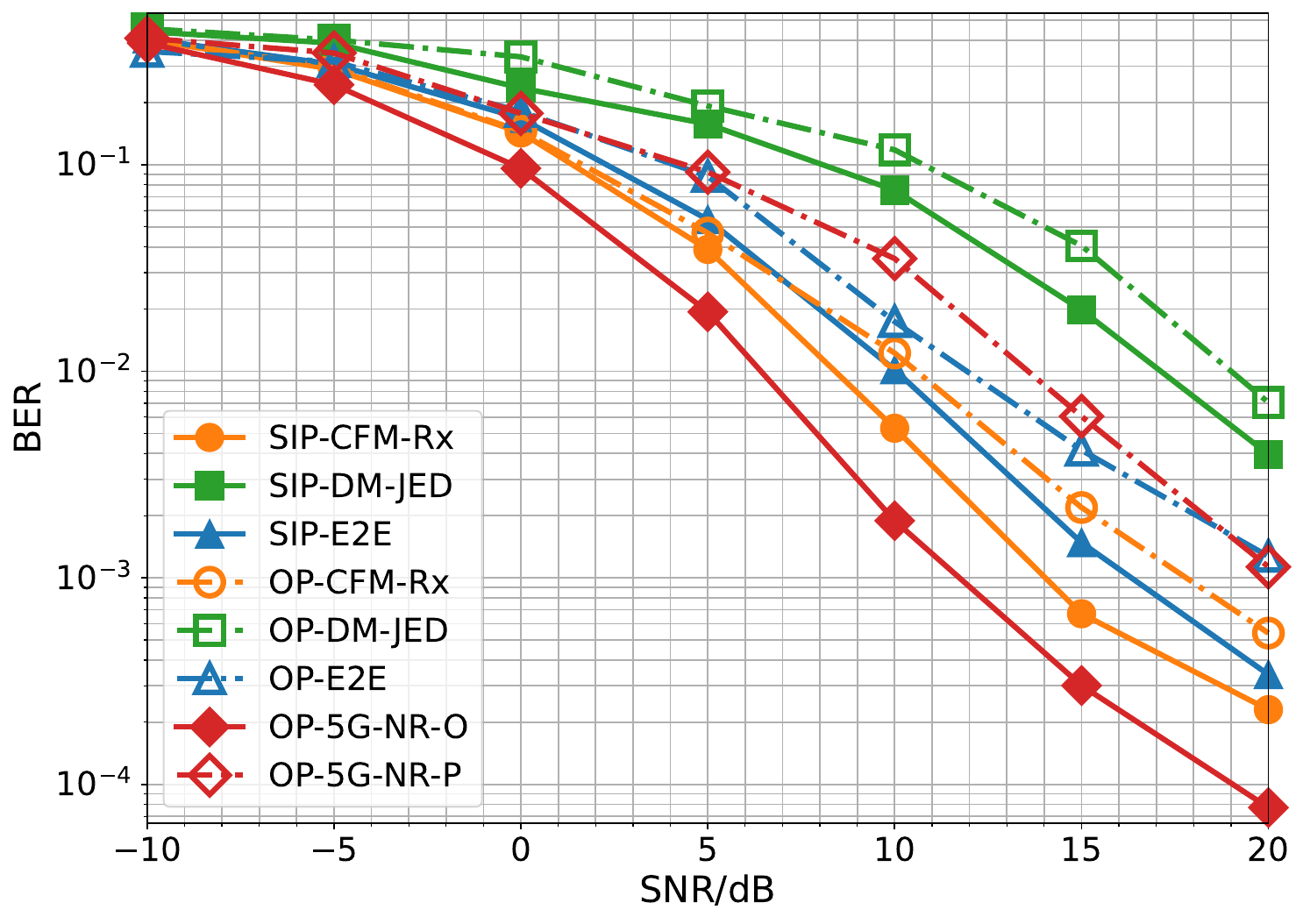}
  \caption{The BER comparisons among various schemes.}
  \label{fig:BER}
\end{figure}
Figure~\ref{fig:BER} presents the uncoded bit error rate (BER) performance across different SNRs. The proposed CFM-Rx demonstrates superior performance to DM-JED and E2E in most cases, thereby demonstrating the significant performance advantage of the CFM-Rx framework. The practical 5G-NR receiver, based on LMMSE channel estimation and equalization, lags behind both the orcale 5G-NR and the proposed CFM-Rx, especially in medium-to-high SNR scenarios, which highlights the sensitivity of model-driven approaches to the accuracy of channel covariance matrices. Across all three schemes, the SIP design yields better BER performance than OP in the vast majority of cases. This suggests that the dense pilot information within most REs of the SIP scheme provides a richer basis for joint estimation and detection, which outweighs the advantage of interference-free pilots in the OP scheme. Furthermore, this BER advantage is coupled with a significant gain in spectral efficiency due to the elimination of pilot overhead, as will be demonstrated next.

\begin{figure}[h]
  \centering
  \includegraphics[width=3.5in,height=2.7in]{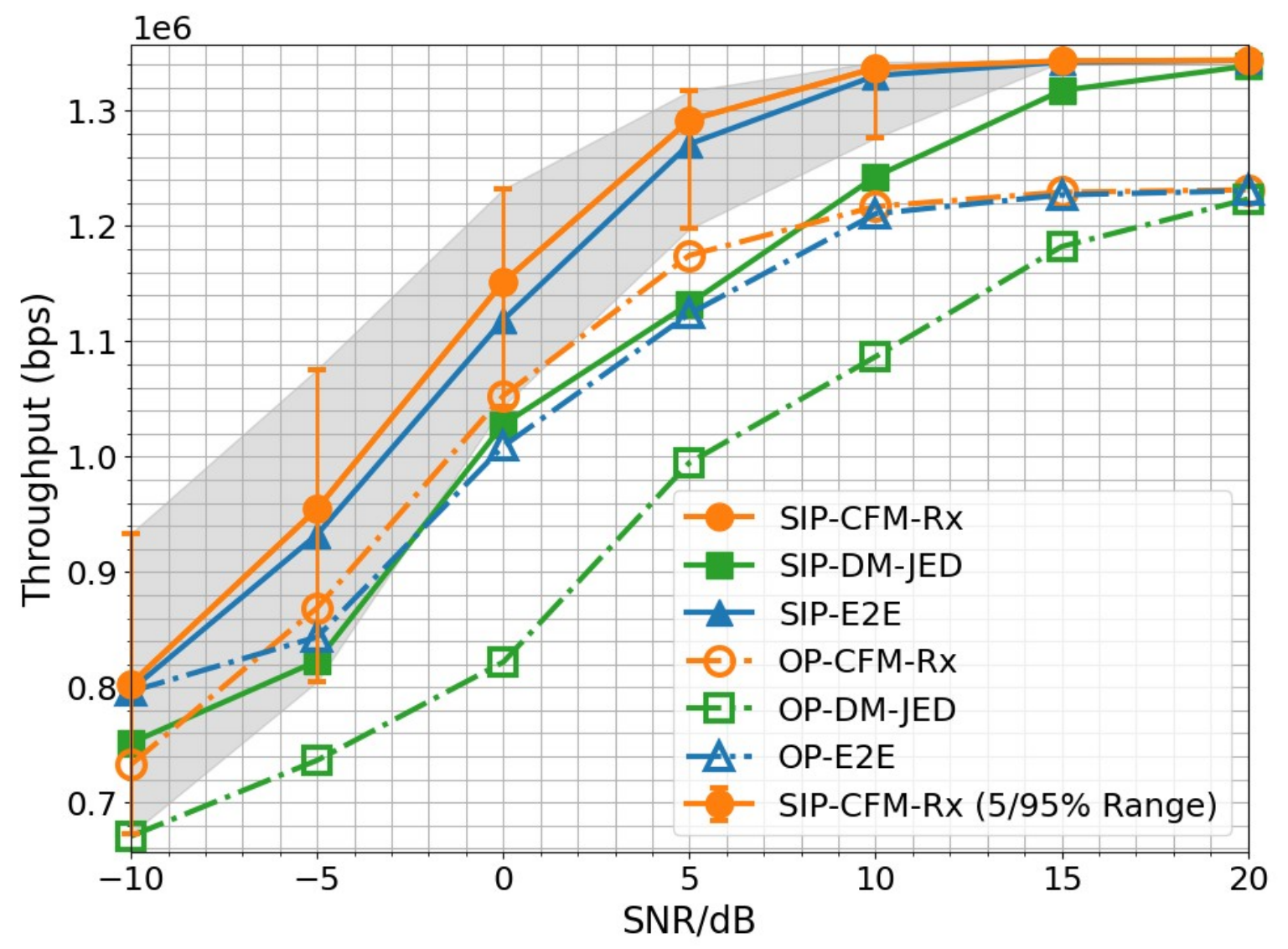}
  \caption{The throughput comparisons among various schemes.}
  \label{fig:TP}
\end{figure}

Figure~\ref{fig:TP} compares the throughput performances of different schemes. The average throughput $R$ (in bits/s) \cite{gu2024learning,aoudia2021end} is given by
\begin{equation}
R = N_{\text{slot}}N_{\text{RE}}\Omega\gamma M(1-\text{BER}), \label{eq:throughput}
\end{equation}
where $N_{\text{slot}}$ is the number of OFDM slots per second, $N_{\text{RE}}$ is the total number of REs per slot, $\Omega$ is the fraction of REs allocated to data, $\gamma$ is the code rate, and $M$ is the number of bits per symbol. It can be seen that all SIP-based methods achieve a higher maximum throughput of about 1.34\,Mbit/s, by utilizing all REs for data transmission. In contrast, OP-based schemes are limited to about 1.23\,Mbit/s, as a fixed fraction of REs is permanently reserved for non-data-bearing pilots. \noindent{However, at low SNR, the throughput for all schemes significantly reduced, indicating degraded performance.} Combining this spectral efficiency with competitive BER performance, CFM-Rx with SIP delivers the highest overall throughput.

\begin{figure}[h]
  \centering
  \includegraphics[width=3.5in,height=2.7in]{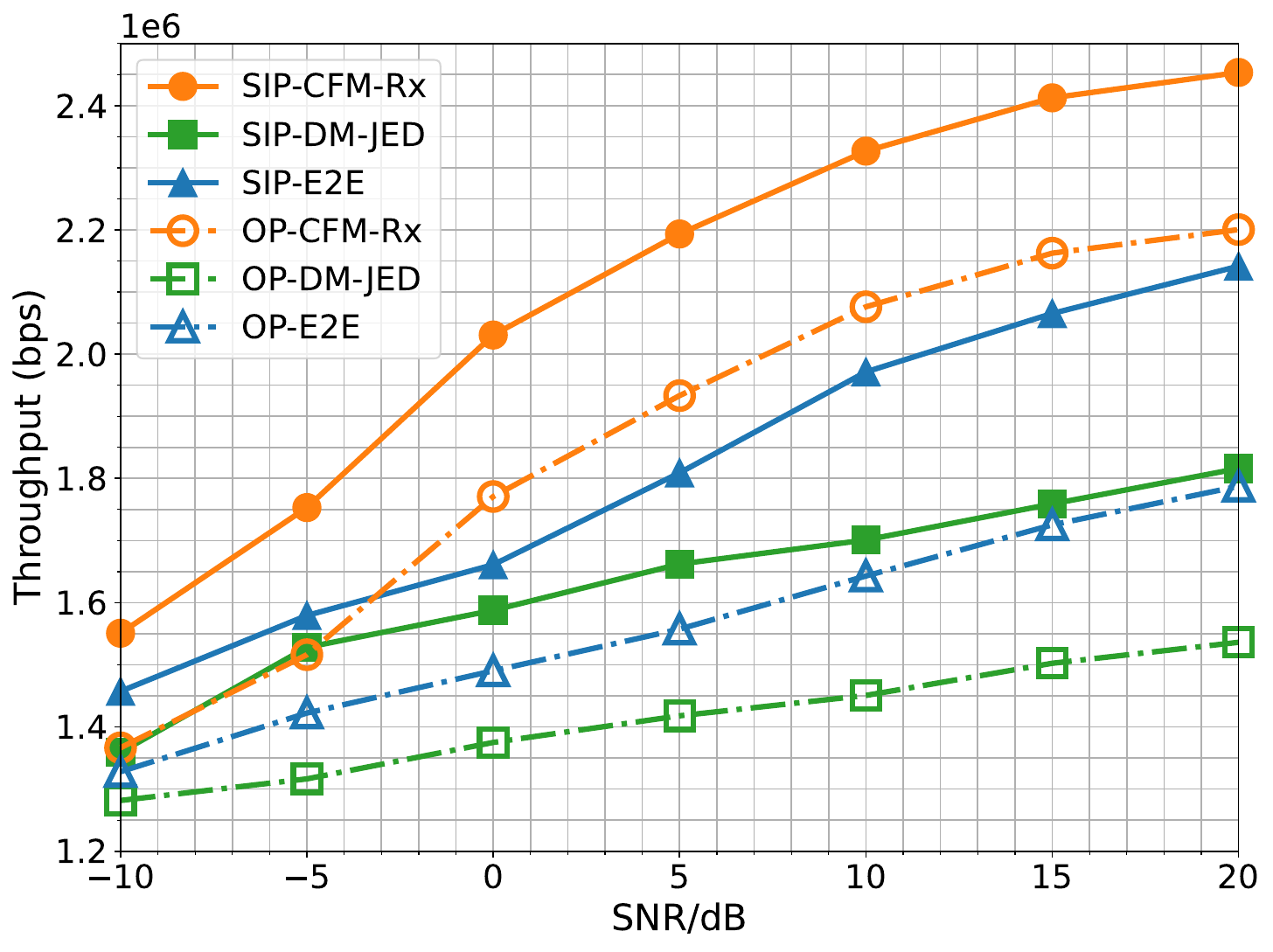}
  \caption{The throughput comparisons among various schemes in a multi-layer ($L=2$) scenario.}
  \label{fig:TP_2L}
\end{figure}

Figure~\ref{fig:TP_2L} evaluates the throughput performance in multi-layer ($L=2$) transmission scenarios. We can observe that even under complex propagation environments with diverse interference and noise, the AI-based receiver demonstrates reliable demodulation capability. In particular, CFM-Rx sustains its performance advantage across both SIP and OP schemes, underscoring its robustness in disentangling signals under intra-stream and inter-stream interference. Moreover, SIP-based schemes consistently outperform their OP counterparts in the evaluated scenarios, as the AI receiver effectively mitigates intra-stream interference introduced by pilot–data overlap, thereby achieving higher throughput despite the presence of superimposed pilots. 

\subsection{Ablation Study}

\begin{figure}[h]
  \centering
  \includegraphics[width=3.5in,height=2.7in]{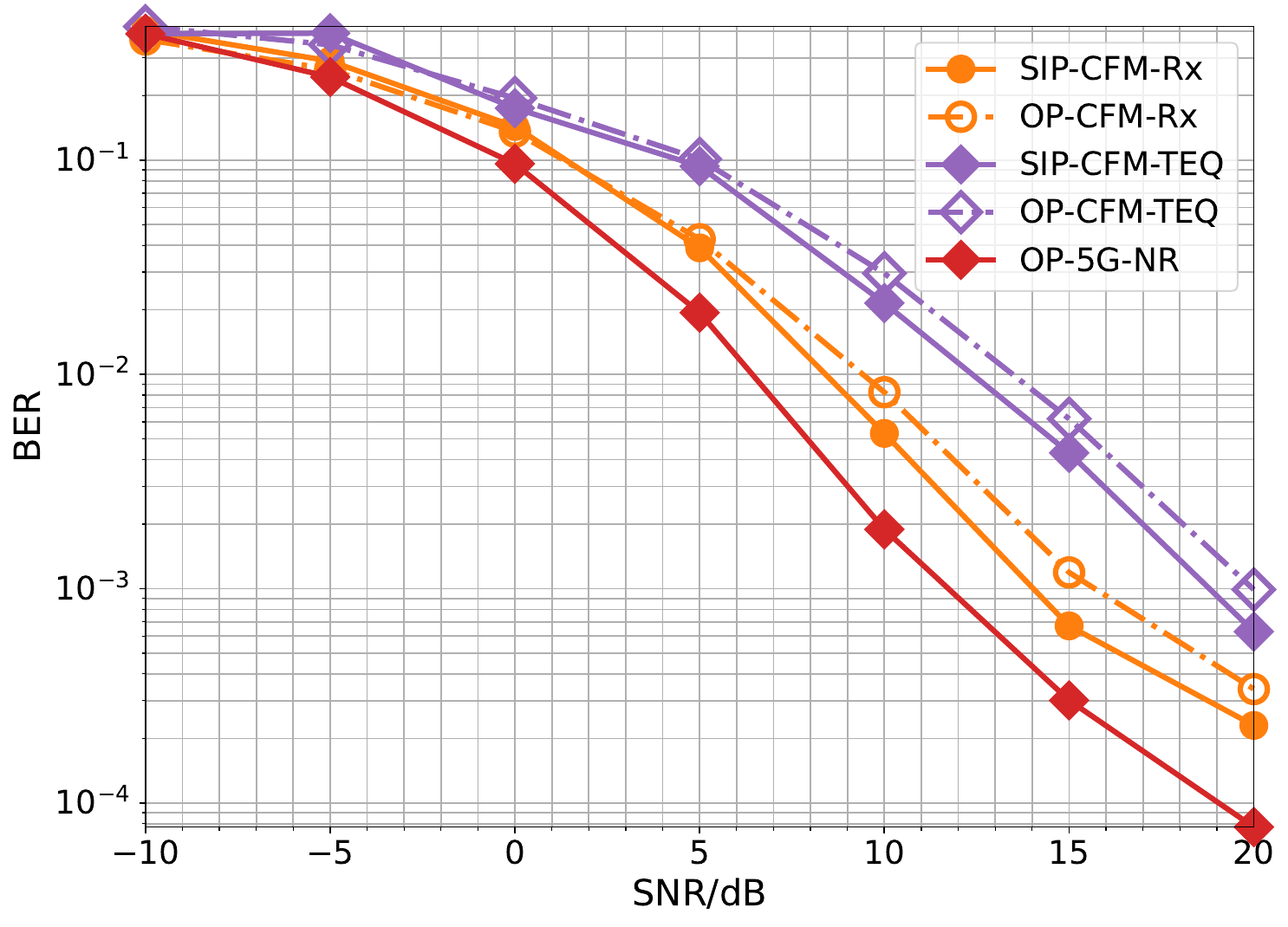}
  \caption{Ablation study: BER performance of CFM-Rx versus CFM-TEQ.}
  \label{fig:BER_ablation}
\end{figure}

Conditional FM–based two-stage equalization (CFM-TEQ) denotes a hybrid scheme in which channel estimation is performed using CFM-Rx, followed by a conventional LMMSE equalizer without mechanisms for correcting residual estimation errors. Owing to this difference in inference structure, Fig.~\ref{fig:BER_ablation} shows that the fully integrated CFM-Rx consistently outperforms CFM-TEQ across the entire SNR range under the considered simulation settings. This gap stems from a fundamental difference in their inference mechanisms. In CFM-TEQ, the LMMSE equalizer operates directly on the channel estimates provided by CFM-Rx but lacks a mechanism to correct residual estimation errors.
As a result, even small deviations in the channel estimates can significantly affect equalization accuracy. In contrast, CFM-Rx performs joint estimation and detection through an iterative refinement process, where the two steps mutually reinforce each other toward convergence. Moreover, as reflected in (\ref{vt_D}), the detection stage in CFM-Rx leverages not only the estimated channel but also the received signal $\mathbf{Y}$ and pilot information $\mathbf{P}$, enabling partial correction of estimation errors and ensuring more reliable symbol recovery. This unified design effectively circumvents the error-propagation problem inherent in cascaded architectures, highlighting the robustness and superiority of CFM-Rx.

\subsection{Convergence Analysis of CFM-Rx}

\begin{figure}[!h]
  \centering
  \includegraphics[width=3.5in,height=2.7in]{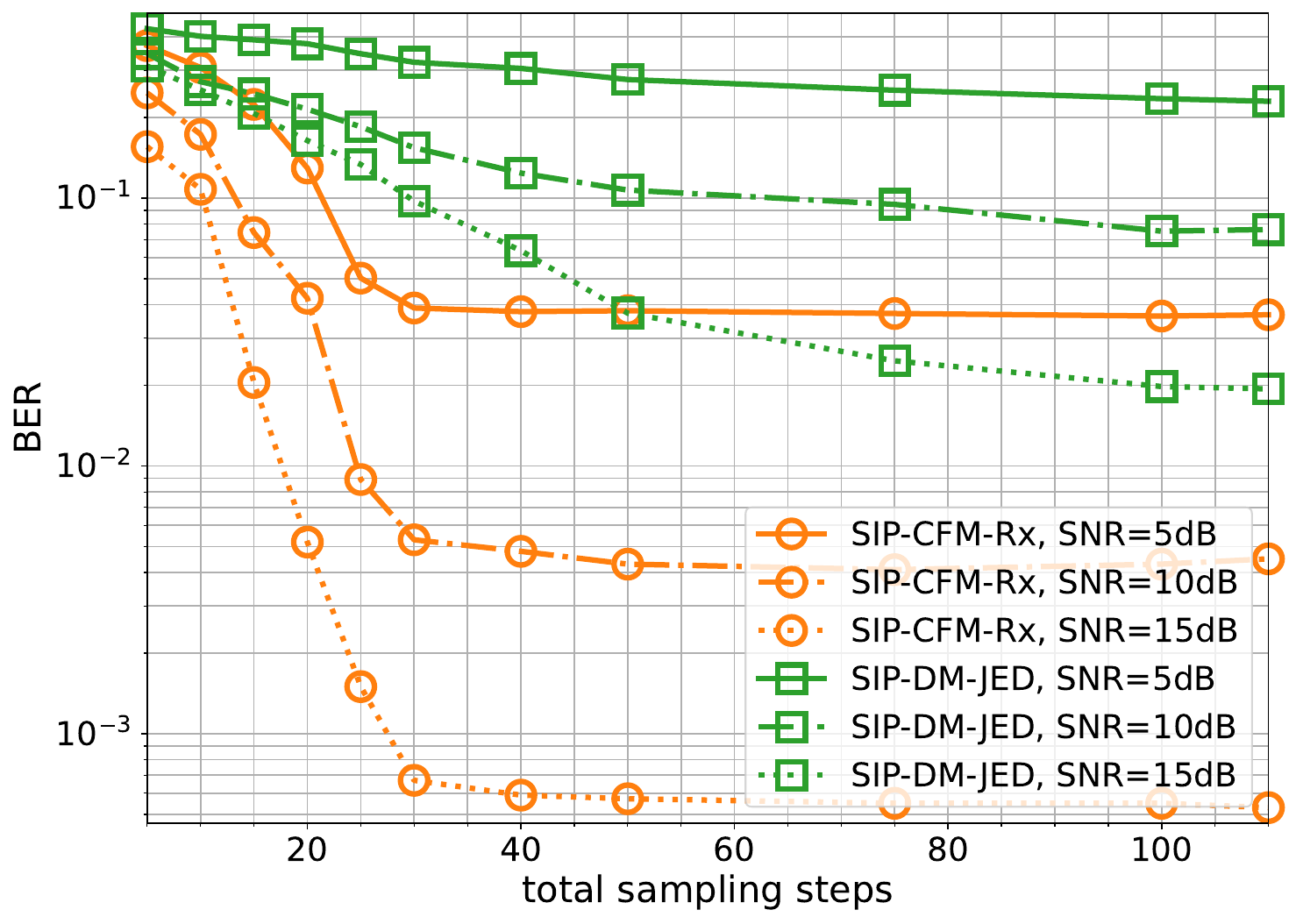}
  \caption{BER performance versus total sampling steps for CFM-Rx and DM-JED.}
  \label{fig:BERwithT}
\end{figure}

\begin{figure}[!h]
  \centering
  \includegraphics[trim=6mm 6mm 6mm 8mm, clip, width=1\columnwidth]{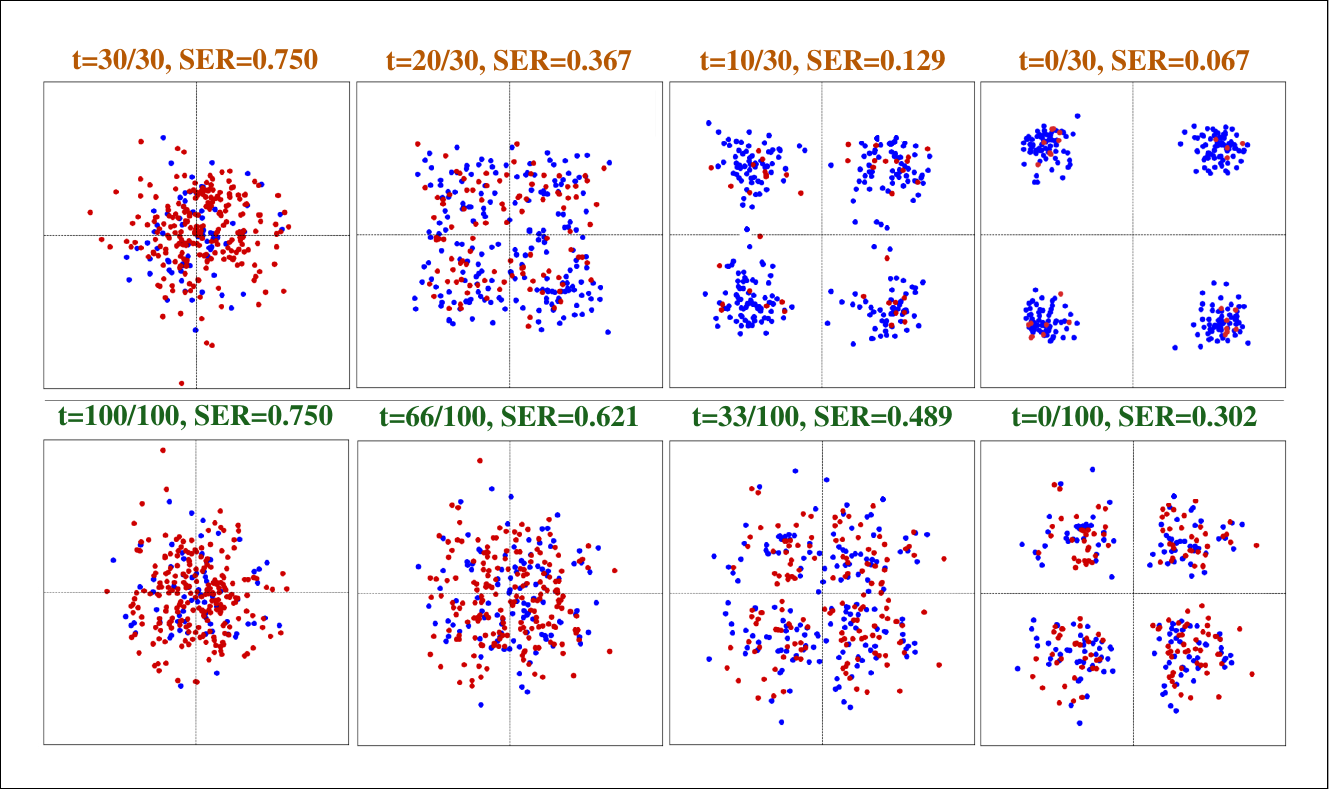}
  \caption{Convergence process comparisons of demodulation symbols in the reverse process. The blue dots represent symbols that fall within the correct decision region, while red dots represent those that would result in a decoding error.}
  \label{fig:scatterswithT}
\end{figure}

Figure~\ref{fig:BERwithT} compares the convergence behavior of CFM-Rx and DM-JED in terms of BER under different SNR levels. CFM-Rx converges within 30 inference steps (each consisting of one predictor-corrector cycle) and then stabilizes, achieving both rapid convergence and high accuracy. In contrast, DM-JED requires more than 100 inference steps to converge, but its final accuracy remains inferior to that of CFM-Rx across all SNR levels. This gap highlights the superior efficiency of CFM-Rx in both convergence speed and precision.

To provide further intuition beyond BER curves, Fig.~\ref{fig:scatterswithT} illustrates the demodulated symbols during the reverse process at SNR=5 dB. The scatter plots reveal distinct clustering dynamics: CFM-Rx rapidly forms tight, well-separated symbol clusters within a few steps, whereas DM-JED produces more dispersed clusters and requires substantially more iterations to achieve partial convergence. These results highlight CFM-Rx's advantages, characterized by faster cluster information and more stable symbol trajectories, in sharp contrast to the stochastic variability inherent in DM-JED.

The root cause of these differences lies in the generative mechanisms. CFM-Rx leverages FM with deterministic ODE sampling, establishing a direct and stable trajectory from the noise distribution to the signal manifold, thereby ensuring rapid and accurate convergence. In contrast, DM-JED employs score-based diffusion with SDE sampling, where stochastic updates introduce randomness, requiring extended iterations to suppress variability and ultimately yielding slower convergence with reduced symbol recovery accuracy.

\subsection{Generalization and Robustness Analysis}

\begin{figure}[h]
  \centering
  \includegraphics[width=3.5in,height=2.7in]{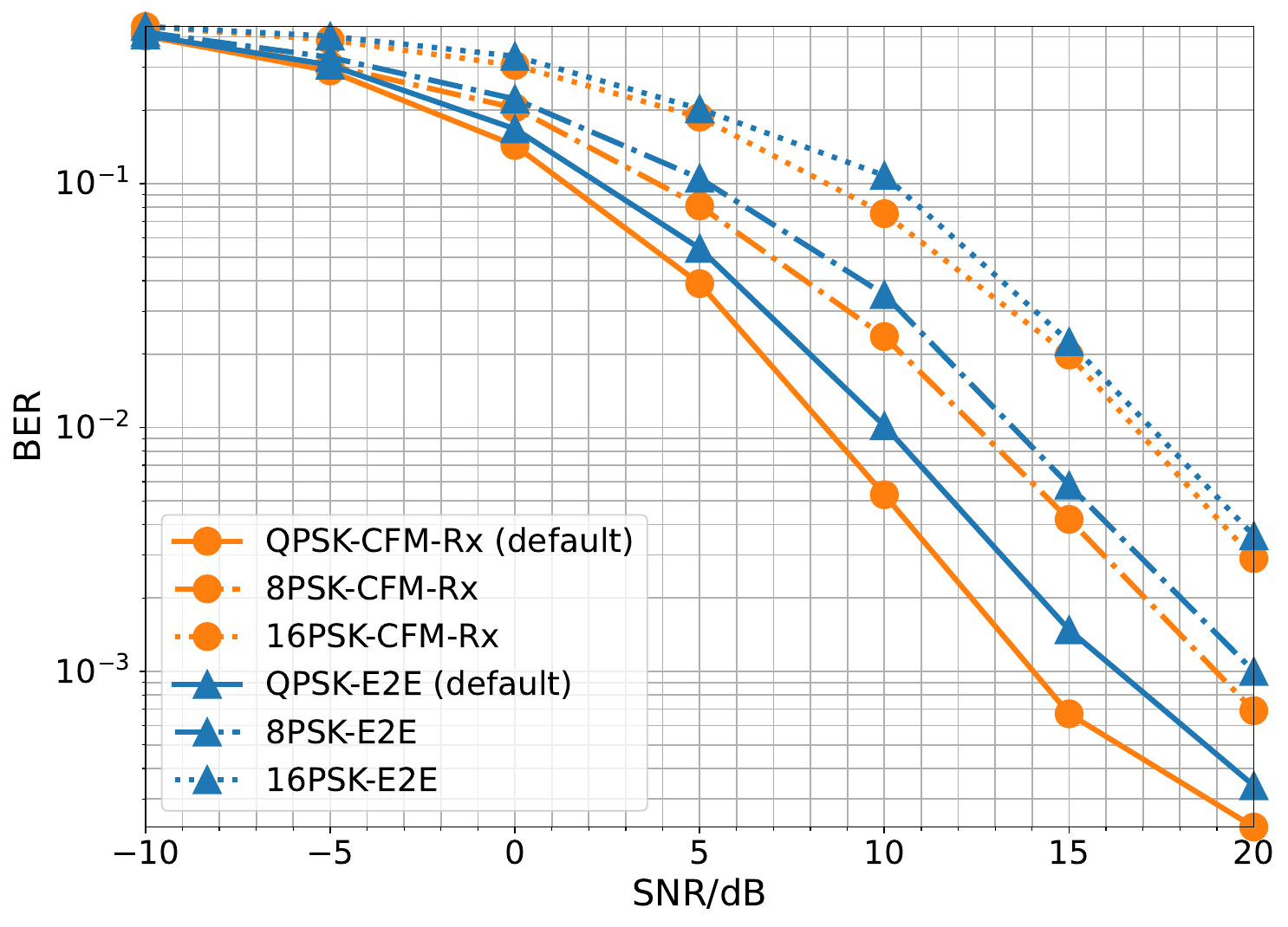}
  \caption{The generalization test across different modulation orders.}
  \label{fig:BER_Generalization}
\end{figure}

Figure~\ref{fig:BER_Generalization} evaluates the BER performance against different modulation orders, including QPSK, 8-PSK, and 16-PSK. CFM-Rx operates with the same network architecture and parameters across all modulation formats, a benefit of its unsupervised training which learns the underlying channel statistics directly. In contrast, the supervised E2E approach is fundamentally constrained, as both its network architecture and input–output mapping are tightly coupled to the modulation order. In our simulations, a dedicated network is trained for each modulation order. This highlights the superior generalization capability of CFM-Rx across different modulation schemes.

Across the simulated SNR range, CFM-Rx consistently outperforms the E2E networks in the evaluated scenarios, each separately trained for a specific modulation, for all tested modulation schemes. These results underscore a key advantage of the unsupervised approach: the learned channel prior naturally generalizes across different modulation formats without requiring any architectural modification or retraining, thereby eliminating the storage and retraining overhead associated with modulation-specific networks. This modulation-agnostic property is particularly advantageous for practical deployment in modern wireless systems that employ adaptive modulation and coding (AMC), where the constellation may change dynamically.

\begin{figure}[h]
  \centering
  \includegraphics[width=3.5in,height=2.7in]{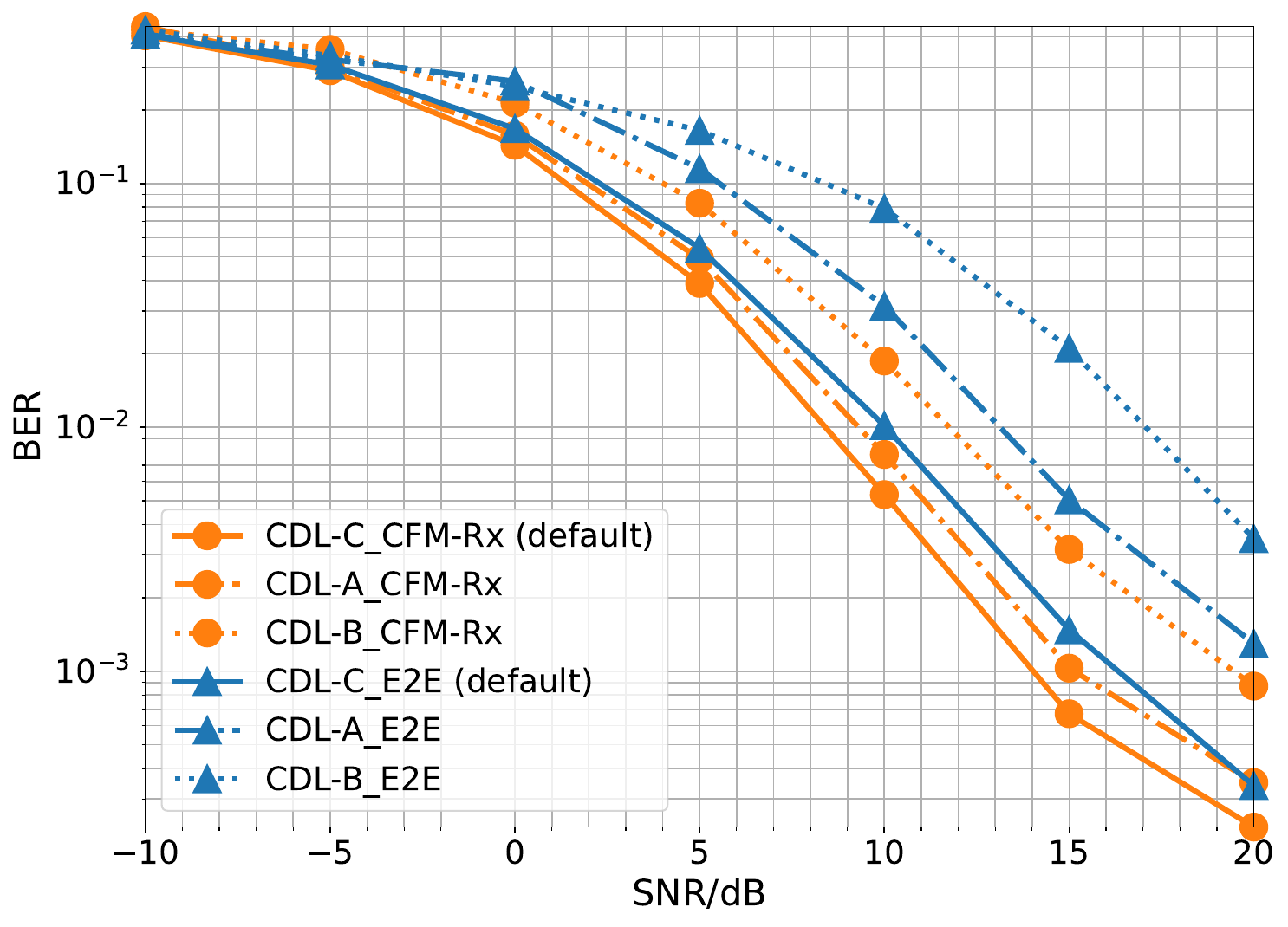}
  \caption{The generalization test across different channel models.}
  \label{fig:BER_Generalization_scenario}
\end{figure}

To assess robustness to distributional shift, we train all models on the CDL-C channel model and evaluate them on two other unseen channel models, namely CDL-A and CDL-B, with BER results shown in Fig.~\ref{fig:BER_Generalization_scenario}. The proposed CFM-Rx generalizes well: when transferred to CDL-A and CDL-B, it incurs only a modest BER increase relative to CDL-C, whereas the E2E suffers a severe performance degradation, particularly on CDL-B scenario. The error ordering \(\text{BER}_{\text{CDL-C}}<\text{BER}_{\text{CDL-A}}<\text{BER}_{\text{CDL-B}}\) aligns with the increasing propagation complexity of these scenarios. This outcome reflects a key limitation of supervised E2E learning: by extracting channel information implicitly from transmit–receive signals, it tends to overfit scenario-specific features, thereby degrading generalization. In contrast, our unsupervised CFM-Rx explicitly guides the model to learn wireless propagation characteristics, leading to a more intrinsic and transferable channel prior and thus superior generalization to complex and previously unseen conditions.

\subsection{Comprehensive analysis}
This section provides a comprehensive comparison between the proposed CFM-Rx and the benchmark methods. We first analyze the algorithmic complexity. For the 5G-NR receiver, the computational burden mainly arises from the channel estimation and equalization modules. Specifically, the LMMSE estimator has a complexity of $\mathcal{O}(N_S N_T N_P^2)$, while LMMSE equalization dominates with $\mathcal{O}(N_S N_T)$, where $N_P$ is the number of pilot REs. This method involves high computational complexity and requires continuously updating the channel covariance estimates, potentially limiting its practical deployment. In contrast, the AI-based receivers in this paper employ CNN architectures. The inference complexity is given by $\mathcal{O}(T k^2 C_{\text{max}}^2 N_S N_T)$, where $T$ is the number of sampling steps, $k$ is the convolutional kernel size, and $C_{\text{max}}$ is the maximum number of channels in the network. The number of sampling steps is $T=1$ for the E2E receiver, $T = 100$ for DM-JED, and $T=30$ for CFM-Rx. The E2E receiver is typically implemented as a convolutional residual neural network \cite{aoudia2021end}. To capture the complex input-output relationships of the receiver system, the E2E receiver requires a large-scale network with increased kernel sizes $k$ and channel dimensions $C_{\text{max}}$, leading to elevated training costs and significant computational overhead. DM-JED uses a U-Net-based architecture similar to CFM-Rx, requires more iterations due to its stochastic updates, leading to higher inference latency. CFM-Rx, on the other hand, uses a compact network design and achieves convergence with fewer iterations, yielding an inference latency slightly higher than E2E but much lower than DM-JED.

Next, we analyze the model parameters, FLOPs, and training dataset requirements. The E2E receiver relies on a large-scale neural network to approximate the sophisticated input-output relationships, requiring $1.315 \times 10^6$ model parameters and 100000 training samples. In contrast, both CFM-Rx and DM-JED learn the underlying wireless channel characteristics directly, requiring only 8000 samples. Their network sizes are considerably smaller, with $1.014 \times 10^5$ and $1.858 \times 10^5$ parameters, respectively. In terms of computational complexity, CFM-Rx demands just $5.705 \times 10^7$ FLOPs, which is lower than DM-JED ($8.942 \times 10^7$ FLOPs) and much smaller than E2E ($7.596 \times 10^8$ FLOPs). This demonstrates its superior efficiency among AI-based receivers. In terms of memory footprint, CFM-Rx is highly parameter-efficient, requiring significantly fewer model parameters, which helps reduce its memory footprint. However, like other deep learning-based methods, it still demands substantial memory for storing intermediate feature maps during inference, particularly when handling large input sizes. In terms of hardware implementation, CFM-Rx benefits from the use of GPUs or specialized accelerators, which may pose challenges for deployment on low-power or embedded devices. On the other hand, traditional LMMSE-based receivers are more straightforward to implement on resource-constrained hardware due to their lower computational and memory demands.

Table \ref{tab:complexity_comparison} summarizes the key metrics, including training category, dataset requirements, parameter scale, FLOPs, generalization, and so on. The results highlight that CFM-Rx offers an optimal trade-off, combining unsupervised training, parameter efficiency, moderate inference steps, and strong generalization.

\begin{table}[h!]
\centering
\caption{Comparison evaluation across complexity, performance, and data requirements.}
\label{tab:complexity_comparison}
\renewcommand{\arraystretch}{1.2}
\resizebox{\columnwidth}{!}{%
\begin{tabular}{l|ccc}
\hline
\textbf{Metric} & \textbf{CFM-Rx} & \textbf{DM-JED} & \textbf{E2E} \\
\hline
Categories      & Unsupervised        & Unsupervised    & Supervised   \\
Training dataset          & Channel data        & Channel data    & Labeled pairs \\
Training sample size    & \textbf{8000}      & 8000      & 100000 \\
Model Parameters       & $\boldsymbol{1.014 \times 10^5}$ & $1.858 \times 10^5$ & $1.315 \times 10^6$ \\
Inference steps  & 30                 & $ 100$        & 1 \\
FLOPs        & $\boldsymbol{5.705 \times 10^7}$ & $8.942 \times 10^7$ & $7.596 \times 10^8$ \\
Generalization         & \textbf{Strong}              & Moderate        & Weak \\
Precision    & \textbf{High}       & Moderate   & High \\
\hline
\end{tabular}}
\end{table}

\section{Conclusion}
This paper proposed CFM-Rx, a novel unsupervised generative receiver for joint channel estimation and data detection in MIMO systems with SIP. By leveraging CFM,  CFM-Rx achieves deterministic, low-latency inference and captures the bidirectional nature of signal propagation, overcoming the reliance on labeled data that limits supervised methods. Its core innovation is a moment-consistent formulation that replaces stochastic SDE sampling with an efficient ODE, ensuring fast and stable inference suitable for physical-layer system. Simulations confirmed that CFM-Rx outperforms existing baselines, closely approaches the LMMSE bound, and achieves superior BER and throughput. Its robust generalization across varying conditions establishes CFM-Rx as an efficient and practical solution for next-generation receivers.

Future work focuses on three specific technical extensions. First, to address the latency challenge, consistency distillation techniques could be investigated to compress the ODE trajectory, enabling high-fidelity generation within a few function evaluations. Second, for high-mobility scenarios, the flow model can be extended to the delay-Doppler domain (e.g., OTFS), explicitly learning the sparse scattering functions to compensate for severe Doppler shifts. Third, the integration of channel coding warrants exploration by developing a soft-output flow mechanism that exchanges log-likelihood ratios (LLRs) with decoders (e.g., LDPC), thereby achieving joint iterative detection and decoding gains.

The proposed framework holds potential for broader applicability in non-orthogonal signaling and integrated sensing and communication (ISAC) scenarios. For instance, in non-orthogonal multiple access (NOMA) schemes, CFM-Rx can effectively mitigate inter-user interference by leveraging its conditional flow matching capabilities. Furthermore, in future ISAC systems for IoT and 6G networks, the framework enables the simultaneous estimation of transmitted data and environmental signals, thereby optimizing resource utilization and system efficiency.


\appendices
\section{Justification of the CFM}
To enable conditional inference under observed context \(\mathbf{z}\), we define a conditional flow ODE driven by the score of the posterior distribution. Specifically, at time \(t\) the conditional velocity field is given by
\begin{equation}
v_t(\mathbf{x}_t | \mathbf{z})
= \frac{\dot{\alpha}_t}{\alpha_t}\mathbf{x}_t
  - \lambda_t\sigma_t\nabla_{\mathbf{x}_t}\log p_t(\mathbf{x}_t| \mathbf{z}).\label{conditionalflowODE}
\end{equation}
To validate this field, we follow the flow matching framework \cite{lipman2023flow,ben2022matching} and show that it satisfies the conditional continuity equation. The following regularity conditions are assumed to hold:
\begin{enumerate}
    \item The joint density \(p_t(\mathbf{x}_t,\mathbf{z})\) is sufficiently smooth with respect to \(\mathbf{x}\).
    \item The boundary behavior satisfies $\lim\limits_{\|\mathbf{x}_t\| \to \infty} p_t(\mathbf{x}_t, \mathbf{z}) = 0, \lim\limits_{\|\mathbf{x}_t\| \to \infty} v_t(\mathbf{x}_t | \mathbf{z}) p_t(\mathbf{x}_t, \mathbf{z}) = 0.$
\end{enumerate}

In what follows, we adopt a unified notation, where \(\mathbf{x}\) denotes the variable of interest and \(\mathbf{z}\) represents the conditioning variables. The conditional dynamics are defined via the ODE $\frac{\mathrm{d}\mathbf{x}_t}{\mathrm{d}t} = v_t(\mathbf{x}_t| \mathbf{z})$. Hence, by Liouville’s theorem discussed in \cite{kardar2007statistical}, the associated joint density \(p_t(\mathbf{x}_t,\mathbf{z})\) satisfies the continuity equation:
\begin{equation}
\partial_t p_t(\mathbf{x}_t,\mathbf{z}) + \nabla_{\mathbf{x}_t}\cdot [v_t(\mathbf{x}_t | \mathbf{z})p_t(\mathbf{x}_t,\mathbf{z})] = 0.\label{continuity}
\end{equation}
Since $\partial_t p_t(\mathbf{z}) = \int \partial_t p_t(\mathbf{x}_t,\mathbf{z})  \mathrm{d}\mathbf{x}_t$, by integrating \eqref{continuity} over \(\mathbf{x}\), we obtain
\begin{equation}
  \partial_t p_t(\mathbf{z}) + \int_{\mathbb{R}^d} \nabla_{\mathbf{x}_t}\cdot \bigl[v_t(\mathbf{x}_t | \mathbf{z})p_t(\mathbf{x}_t,\mathbf{z})\bigr] \mathrm{d}\mathbf{x}_t = 0,
  \label{ptz}
\end{equation}
where $\mathbb{R}^d$ denotes the $d$-dimensional Euclidean space. Applying the divergence theorem to the second term yields a surface integral over a sphere at infinity. Under Condition 2), this integral vanishes, and it follows that \(\partial_t p_t(\mathbf{z})=0\). The time derivative of the conditional density $p_t(\mathbf{x}_t | \mathbf{z})$ is given by
\begin{equation}
  \partial_t p_t(\mathbf{x}_t | \mathbf{z}) = \frac{\partial_t p_t(\mathbf{x}_t,\mathbf{z})}{p_t(\mathbf{z})} - \frac{p_t(\mathbf{x}_t,\mathbf{z})}{p_t(\mathbf{z})^2} \partial_t p_t(\mathbf{z}).\label{ptxz}
\end{equation}
Substituting \(\partial_t p_t(\mathbf{z})=0\) into (\ref{ptxz}), yields \(\partial_t p_t(\mathbf{x}_t | \mathbf{z}) = \partial_t p_t(\mathbf{x}_t,\mathbf{z})/p_t(\mathbf{z})\). Finally, dividing \eqref{continuity} by \(p_t(\mathbf{z})\), we arrive at the conditional continuity equation:
\[
\partial_t p_t(\mathbf{x}_t | \mathbf{z}) + \nabla_{\mathbf{x}_t} \cdot [v_t(\mathbf{x}_t| \mathbf{z})p_t(\mathbf{x}_t | \mathbf{z})] = 0.
\]
Since our proposed conditional velocity field satisfies this equation, it is guaranteed to generate a valid probability flow that correctly transports the conditional distribution $p_t(\mathbf{x}_t| \mathbf{z})$ over time, thus establishing the theoretical soundness of our approach.

\section*{REFERENCES}

\def\refname{\vadjust{\vspace*{-1em}}} 


\begin{thebibliography}{00}

\bibitem{mimo1}
H. Q. Ngo, E. G. Larsson, and T. L. Marzetta, “Energy and spectral efficiency of very large multiuser MIMO systems,” \textit{IEEE Trans. Commun.}, vol. 61, no. 4, pp. 1436–1449, Apr. 2013.

\bibitem{mimo2}
T. L. Marzetta, “Noncooperative cellular wireless with unlimited numbers of base station antennas,” \textit{IEEE Trans. Wireless Commun.}, vol. 9, no. 11, pp. 3590–3600, Nov. 2010.

\bibitem{chi2011}
Y. Chi, L. L. Scharf, A. Pezeshki, and A. R. Calderbank, “Sensitivity to basis mismatch in compressed sensing,” \textit{IEEE Trans. Signal Process.}, vol. 59, no. 5, pp. 2182–2195, May 2011.

\bibitem{li2024channel}
Y. Li, G. Li, Z. Wen, S. Han, S. Gao, et al., “Channel modeling aided dataset generation for AI-enabled CSI feedback: advances, challenges, and solutions,” \textit{IEEE Commun. Stand. Mag.}, vol. 8, no. 4, pp. 72-78, Dec. 2024.

\bibitem{kim2008map}
J. G. Kim and J. T. Lim, “Map-based channel estimation for MIMO–OFDM over fast Rayleigh fading channels,” \textit{IEEE Trans. Veh. Technol.}, vol. 57, no. 3, pp. 1963–1968, May 2008.

\bibitem{rag2005improving}
M. R. Raghavendra and K. Giridhar, “Improving channel estimation in OFDM systems for sparse multipath channels,” \textit{IEEE Signal Process. Lett.}, vol. 12, no. 1, pp. 52–55, Jan. 2005.

\bibitem{bajwa2010com}
W. U. Bajwa, J. Haupt, A. M. Sayeed, and R. Nowak, “Compressed channel sensing: A new approach to estimating sparse multipath channels,” \textit{Proc. IEEE}, vol. 98, no. 6, pp. 1058–1076, Jun. 2010.

\bibitem{scaling}
F. Rusek, D. Persson, B. K. Lau, E. G. Larsson, T. L. Marzetta, O. Edfors, and F. Tufvesson, “Scaling up MIMO: Opportunities and challenges with very large arrays,” \textit{IEEE Signal Process. Mag.}, vol. 30, no. 1, pp. 40–60, Jan. 2013.


\bibitem{zilberstein2024diffusion}
N. Zilberstein, A. Swami, and S. Segarra, “Joint channel estimation and data detection in massive MIMO systems based on diffusion models,” in \textit{Proc. IEEE Int. Conf. Acoust., Speech Signal Process. (ICASSP)}, pp. 13291–13295, 2024.

\bibitem{bjornson2017massive}
E. Björnson, J. Hoydis, and L. Sanguinetti, 
"Massive MIMO networks: Spectral, energy, and hardware efficiency," \textit{Found. Trends Signal Process.}, vol. 11, no. 3–4, pp. 154–655, 2017.

\bibitem{adhikary2013}
A. Adhikary, J. Nam, J.-Y. Ahn, and G. Caire, “Joint spatial division and multiplexing—the large-scale array regime,” \textit{IEEE Trans. Inf. Theory}, vol. 59, no. 10, pp. 6441–6463, Oct. 2013.

\bibitem{xie2024cellfree}
M. Xie, H. Smith, T. Johnson, Y. Zhao, and A. Garcia,  
“Superimposed pilots for cell-free massive MIMO over spatially-correlated rician fading channels,”  
\textit{IEEE Trans. Wirel. Commun.}, vol. 23, no. 5, pp. 3312–3328, 2024.

\bibitem{gupta2024affine}
A. Gupta, M. Jafri, S. Srivastava, A. K. Jagannatham, and L. Hanzo,  
“An affine precoded superimposed pilot-based mmWave MIMO-OFDM ISAC system,” \textit{IEEE Open J. Commun. Soc.}, vol. 5, pp. 1504–1524, 2024.

\bibitem{gu2024learning}
R. Gu, J. Xu, C. Qian, W. Xu, and R. Xie,  
“Learning power allocation and channel estimation for superimposed pilot-assisted multiuser MIMO,” \textit{IEEE Wirel. Commun. Lett.}, vol. 13, no. 11, pp. 3025–3029, 2024.

\bibitem{ashikhmin2012pilots}
A. Ashikhmin and T. L. Marzetta, “Pilot contamination precoding in multi-cell large scale antenna systems,” in \textit{Proc. IEEE Int. Symp. Inf. Theory (ISIT)}, pp. 1137–1141, Jul. 2012.

\bibitem{li2025learning}
X. Li, X. Zhou, Y. Cao, J. Zhang, C.-K. Wen, S. Jin, and X. Li, “Learning-aided iterative receiver for superimposed pilots in MIMO-OFDM systems,”  
\textit{arXiv preprint:2507.10074}, 2025.

\bibitem{li2025iterative}
X. Li, X. Zhou, J. Zhang, C.-K. Wen, and S. Jin,  
“AI-driven iterative receiver for superimposed pilot schemes in MIMO-OFDM systems,” in \textit{Proc. IEEE Wireless Commun. Netw. Conf. (WCNC)}, pp. 1–6, 2025.

\bibitem{aoudia2021end}
F. A. Aoudia and J. Hoydis, “End-to-end learning for OFDM: From neural receivers to pilotless communication,” \textit{IEEE Trans. Wireless Commun.}, vol. 21, no. 2, pp. 1049–1063, Feb. 2022.

\bibitem{neumann2018learning}
D. Neumann, T. Wiese, and W. Utschick, “Learning the MMSE channel estimator,” \textit{IEEE Trans. Signal Process.}, vol. 66, no. 11, pp. 2905–2917, Jun. 2018.

\bibitem{zou2025deeppilot}
J. Zou, J. Xiao, Q. Mao, S. Liu, B. Xiao, and Y. Liang,
“Deep receiver for multi-layer data transmission with superimposed pilots,” in \textit{Proc. IEEE Int. Conf. Acoust., Speech Signal Process. (ICASSP)}, 2025, pp. 1–5.

\bibitem{soltani2019deep}
M. Soltani, V. Pourahmadi, A. Mirzaei, and H. Sheikhzadeh, “Deep learning-based channel estimation,” \textit{IEEE Commun. Lett.}, vol. 23, no. 4, pp. 652–655, Apr. 2019.

\bibitem{savaux2017lmmse}
V. Savaux and Y. Louët, “LMMSE channel estimation in OFDM context: a review,” \textit{IET Signal Process.}, vol. 11, no. 2, pp. 123–134, 2017.

\bibitem{unsup2}
S. Dorner, S. Cammerer, J. Hoydis, and S. ten Brink, “Deep learning based communication over the air,” \textit{IEEE J. Sel. Topics Signal Process.}, vol. 12, no. 1, pp. 132–143, Feb. 2018.

\bibitem{song2021score}
Y. Song, J. Sohl-Dickstein, D. P. Kingma, A. Kumar, S. Ermon, and B. Poole, “Score-based generative modeling through stochastic differential equations,” in \textit{Proc. Int. Conf. Learn. Representations (ICLR)}, 2021.

\bibitem{ho2020denoising}
J. Ho, A. Jain, and P. Abbeel, 
“Denoising diffusion probabilistic models,” in \textit{Proc. Advances in Neural Inf. Process. Syst. (NeurIPS)}, 
Vancouver, BC, Canada, pp. 6840–6851, Dec. 2020.

\bibitem{song2019generative}
Y. Song and S. Ermon,
“Generative modeling by estimating gradients of the data distribution,” 
\textit{Advances in Neural Information Processing Systems (NeurIPS)}, vol. 32, pp. 11895–11907, Dec. 2019.

\bibitem{yudiffusion}
Y. Li, R. Zhang, Y. Liu, C. Shao, J. Jin, et al., “Denoising and augmentation: A dual use of diffusion model for enhanced CSI recovery,” in \textit{Int. Conf. Wireless Commun. Signal Process (WCSP)}, 2025.

\bibitem{zilberstein2024joint}
N. Zilberstein, A. Swami, and S. Segarra,  
"Joint channel estimation and data detection in massive MIMO systems based on diffusion models,"  
in \textit{Proc. IEEE Int. Conf. Acoust., Speech Signal Process. (ICASSP)}, pp. 13291–13295, 2024.

\bibitem{214}
3GPP TS 38.214, v19.2.0, ``NR; Physical layer procedures for data,'' Dec. 2025.

\bibitem{lipman2023flow}
Y. Lipman, T. R. Huang, and D. B. Dunson, “Flow matching for generative modeling,” in \textit{Proc. Adv. Neural Inf. Process. Syst. (NeurIPS)}, 2023.

\bibitem{ben2022matching}
H. Ben-Hamu, S. Cohen, J. Bose, B. Amos, et al., "Matching normalizing flows and probability paths on manifolds," \textit{arXiv preprint arXiv:2207.04711}, 2022.

\bibitem{kobyzev2020normalizing}
I. Kobyzev, S. Prince, and M. A. Brubaker, “Normalizing flows: An introduction and review of current methods,” \textit{IEEE Trans. Pattern Anal. Mach. Intell.}, vol. 43, no. 11, pp. 3964–3979, Nov. 2021.

\bibitem{papamakarios2019normalizing}
G. Papamakarios, E. Nalisnick, D. J. Rezende, S. Mohamed, and B. Lakshminarayanan, 
"Normalizing flows for probabilistic modeling and inference," 
\textit{J. Mach. Learn. Res.}, vol. 22, no. 57, pp. 1–64, 2019.

\bibitem{kingma2018glow}
D. P. Kingma and P. Dhariwal, 
"Glow: Generative flow with invertible $1{\times}1$ convolutions," 
in \textit{Adv. Neural Inf. Process. Syst. (NeurIPS)}, 2018.

\bibitem{jiang2026recursive}
Z. Jiang, F. Zhu, C. Huang, R. Jin, Z. Yang, X. Chen, Z. Zhang, and M. Debbah,
“Recursive flow: A generative framework for MIMO channel estimation,”
\textit{arXiv preprint arXiv:2601.15767}, Jan. 2026.

\bibitem{wang1}
H. Zhu and J. Wang, “Chunk-based resource allocation in OFDMA systems - Part I: chunk allocation,” \textit{IEEE Trans. on Commun.}, vol. 57, no. 9, pp. 2734-2744, Sept. 2009.

\bibitem{wang2}
H. Zhu and J. Wang, “Chunk-based resource allocation in OFDMA systems - Part II: joint chunk, power and bit allocation,” \textit{IEEE Trans. on Commun.}, vol. 60, no. 2, pp. 499-509, Feb. 2012.


\bibitem{chen2023probability}
S. Chen, S. Chewi, H. Lee, L. Mackey, and A. Risteski, 
“The probability flow ODE is provably fast,” 
in \textit{Adv. Neural Inf. Process. Syst.}, vol. 36, pp. 68552–68575, 2023.

\bibitem{chen2018neural}
R. T. Q. Chen, Y. Rubanova, J. Bettencourt, and D. K. Duvenaud, 
“Neural ordinary differential equations,” 
in \textit{Adv. Neural Inf. Process. Syst.}, vol. 31, pp. 6572–6583, 2018.

\bibitem{mccann1997convexity}
R. J. McCann, “A convexity principle for interacting gases,” \textit{Adv. Math.}, vol. 128, no. 1, pp. 153–179, 1997.


\bibitem{kardar2007statistical}
M. Kardar, "Statistical physics of particles," \textit{Cambridge University Press}, Cambridge, UK, 2007.

\bibitem{efron2011tweedie}
B. Efron, "Tweedie's formula and selection bias," 
\textit{J. Amer. Stat. Assoc.}, vol. 106, no. 496, pp. 1602–1614, 2011.

\bibitem{blei2017variational}
D. M. Blei, A. Kucukelbir, and J. D. McAuliffe, "Variational inference: A review for statisticians,"
\textit{J. of the Amer. Stat. Assoc.}, vol. 112, no. 518, pp. 859–877, 2017.


\bibitem{leimkuhler2015molecular}
B. Leimkuhler and C. Matthews, “Molecular dynamics: With deterministic and stochastic numerical methods,” \textit{Interdisciplinary Appl. Math.} Springer, vol. 39, 2015.

\bibitem{hairer1993solving}
E. Hairer, S. P. N{\o}rsett, and G. Wanner, 
“Solving ordinary differential equations I: Nonstiff problems,” 
2nd ed., Springer Verlag, Berlin, 1993.

\bibitem{901}
3GPP TR 38.901, v17.0.0, ``Study on channel model for frequencies from 0.5 to 100 GH,'' 2022.

\bibitem{noh2006low}
M. Noh, Y. Lee, and H. Park, “Low complexity LMMSE channel estimation for OFDM,” \textit{IEE Proc. Commun.}, vol. 153, no. 5, pp. 645–650, 2006.

\bibitem{jiang2011performance}
Y. Jiang, M. K. Varanasi, and J. Li, “Performance analysis of ZF and MMSE equalizers for MIMO systems: An in-depth study of the high SNR regime,” \textit{IEEE Trans. Inf. Theory}, vol. 57, no. 4, pp. 2008–2026, Apr. 2011.

\end{thebibliography}
\end{document}